\documentclass[%
reprint,
superscriptaddress,
nofootinbib,
nobibnotes,
 amsmath,
 amssymb,
 mathtools,
 aps,
 prl,
twocolumn,
]{revtex4-2}

\usepackage{float}
\usepackage[permil]{overpic}
\usepackage{graphicx, comment}
\usepackage{dcolumn}
\usepackage{bm}
\usepackage{xcolor}
\usepackage{hyperref}
\usepackage[mathlines]{lineno}
\usepackage{mathtools}
\usepackage{amsmath}
\usepackage[normalem]{ulem}
\usepackage{amsfonts}
\usepackage{url}
\usepackage{hyphenat}
\newcommand{\nue}{$\nu_{e}$}
\newcommand{\numu}{$\nu_{\mu}$}
\newcommand{\pio}{$\pi^0$}
\newcommand\brabarb{\scalebox{.3}{(}\raisebox{-1.7pt}[0pt][0pt]{--}\scalebox{.3}{)}}

\begin{document}
\title{First constraints on light sterile neutrino oscillations from combined appearance and disappearance searches with the MicroBooNE detector}

\newcommand{\Bern}{Universit{\"a}t Bern, Bern CH-3012, Switzerland}
\newcommand{\BNL}{Brookhaven National Laboratory (BNL), Upton, NY, 11973, USA}
\newcommand{\UCSB}{University of California, Santa Barbara, CA, 93106, USA}
\newcommand{\Cambridge}{University of Cambridge, Cambridge CB3 0HE, United Kingdom}
\newcommand{\CIEMAT}{Centro de Investigaciones Energ\'{e}ticas, Medioambientales y Tecnol\'{o}gicas (CIEMAT), Madrid E-28040, Spain}
\newcommand{\Chicago}{University of Chicago, Chicago, IL, 60637, USA}
\newcommand{\Cincinnati}{University of Cincinnati, Cincinnati, OH, 45221, USA}
\newcommand{\CSU}{Colorado State University, Fort Collins, CO, 80523, USA}
\newcommand{\Columbia}{Columbia University, New York, NY, 10027, USA}
\newcommand{\Edinburgh}{University of Edinburgh, Edinburgh EH9 3FD, United Kingdom}
\newcommand{\FNAL}{Fermi National Accelerator Laboratory (FNAL), Batavia, IL 60510, USA}
\newcommand{\Granada}{Universidad de Granada, Granada E-18071, Spain}
\newcommand{\Harvard}{Harvard University, Cambridge, MA 02138, USA}
\newcommand{\IIT}{Illinois Institute of Technology (IIT), Chicago, IL 60616, USA}
\newcommand{\KSU}{Kansas State University (KSU), Manhattan, KS, 66506, USA}
\newcommand{\Lancaster}{Lancaster University, Lancaster LA1 4YW, United Kingdom}
\newcommand{\LANL}{Los Alamos National Laboratory (LANL), Los Alamos, NM, 87545, USA}
\newcommand{\Louisiana}{Louisiana State University, Baton Rouge, LA, 70803, USA}
\newcommand{\Manchester}{The University of Manchester, Manchester M13 9PL, United Kingdom}
\newcommand{\MIT}{Massachusetts Institute of Technology (MIT), Cambridge, MA, 02139, USA}
\newcommand{\Michigan}{University of Michigan, Ann Arbor, MI, 48109, USA}
\newcommand{\Minnesota}{University of Minnesota, Minneapolis, MN, 55455, USA}
\newcommand{\NMSU}{New Mexico State University (NMSU), Las Cruces, NM, 88003, USA}
\newcommand{\Oxford}{University of Oxford, Oxford OX1 3RH, United Kingdom}
\newcommand{\Pitt}{University of Pittsburgh, Pittsburgh, PA, 15260, USA}
\newcommand{\Rutgers}{Rutgers University, Piscataway, NJ, 08854, USA}
\newcommand{\SLAC}{SLAC National Accelerator Laboratory, Menlo Park, CA, 94025, USA}
\newcommand{\SDSMT}{South Dakota School of Mines and Technology (SDSMT), Rapid City, SD, 57701, USA}
\newcommand{\Maine}{University of Southern Maine, Portland, ME, 04104, USA}
\newcommand{\Syracuse}{Syracuse University, Syracuse, NY, 13244, USA}
\newcommand{\TelAviv}{Tel Aviv University, Tel Aviv, Israel, 69978}
\newcommand{\Tennessee}{University of Tennessee, Knoxville, TN, 37996, USA}
\newcommand{\UTA}{University of Texas, Arlington, TX, 76019, USA}
\newcommand{\Tufts}{Tufts University, Medford, MA, 02155, USA}
\newcommand{\VTech}{Center for Neutrino Physics, Virginia Tech, Blacksburg, VA, 24061, USA}
\newcommand{\Warwick}{University of Warwick, Coventry CV4 7AL, United Kingdom}
\newcommand{\Yale}{Wright Laboratory, Department of Physics, Yale University, New Haven, CT, 06520, USA}

\affiliation{\Bern}
\affiliation{\BNL}
\affiliation{\UCSB}
\affiliation{\Cambridge}
\affiliation{\CIEMAT}
\affiliation{\Chicago}
\affiliation{\Cincinnati}
\affiliation{\CSU}
\affiliation{\Columbia}
\affiliation{\Edinburgh}
\affiliation{\FNAL}
\affiliation{\Granada}
\affiliation{\Harvard}
\affiliation{\IIT}
\affiliation{\KSU}
\affiliation{\Lancaster}
\affiliation{\LANL}
\affiliation{\Louisiana}
\affiliation{\Manchester}
\affiliation{\MIT}
\affiliation{\Michigan}
\affiliation{\Minnesota}
\affiliation{\NMSU}
\affiliation{\Oxford}
\affiliation{\Pitt}
\affiliation{\Rutgers}
\affiliation{\SLAC}
\affiliation{\SDSMT}
\affiliation{\Maine}
\affiliation{\Syracuse}
\affiliation{\TelAviv}
\affiliation{\Tennessee}
\affiliation{\UTA}
\affiliation{\Tufts}
\affiliation{\VTech}
\affiliation{\Warwick}
\affiliation{\Yale}

\author{P.~Abratenko} \affiliation{\Tufts}
\author{D.~Andrade~Aldana} \affiliation{\IIT}
\author{J.~Anthony} \affiliation{\Cambridge}
\author{L.~Arellano} \affiliation{\Manchester}
\author{J.~Asaadi} \affiliation{\UTA}
\author{A.~Ashkenazi}\affiliation{\TelAviv}
\author{S.~Balasubramanian}\affiliation{\FNAL}
\author{B.~Baller} \affiliation{\FNAL}
\author{G.~Barr} \affiliation{\Oxford}
\author{J.~Barrow} \affiliation{\MIT}\affiliation{\TelAviv}
% ...

\collaboration{The MicroBooNE Collaboration}
\thanks{microboone\_info@fnal.gov}\noaffiliation

\date{\today}

\begin{abstract}
We present a search for eV-scale sterile neutrino oscillations in the MicroBooNE liquid argon detector, simultaneously considering all possible appearance and disappearance effects within the $3+1$ active-to-sterile neutrino oscillation framework. We analyze the neutrino candidate events for the recent measurements of charged-current \nue\ and \numu\ interactions in the MicroBooNE detector, using data corresponding to an exposure of 6.37$\times$10$^{20}$ protons on target from the Fermilab booster neutrino beam.
We observe no evidence of light sterile neutrino oscillations and derive exclusion contours at the $95\%$ confidence level in the plane of the mass-squared splitting $\Delta m^2_{41}$ and the sterile neutrino mixing angles $\theta_{\mu e}$ and $\theta_{ee}$, excluding part of the parameter space allowed by experimental anomalies.
Cancellation of \nue\ appearance and \nue\ disappearance effects due to the full $3+1$ treatment of the analysis leads to a degeneracy when determining the oscillation parameters, which is discussed in this paper and will be addressed by future analyses.
\end{abstract}

\maketitle

The discoveries of solar~\cite{SNO:2001kpb} and atmospheric neutrino oscillations~\cite{Super-Kamiokande:1998kpq} have motivated a broad experimental program dedicated to studying neutrino mixing. While most measurements~\cite{SNO:2011hxd, Super-Kamiokande:2017yvm, IceCube:2017lak, KamLAND:2013rgu, DayaBay:2018yms, RENO:2018dro, DoubleChooz:2015mfm, T2K:2021xwb, NOvA:2021nfi, MINOS:2020llm, OPERA:2018nar} are consistent with three-flavor ($3\nu$) neutrino oscillations as described by the Pontecorvo-Maki-Nakagawa-Sakata (PMNS) formalism~\cite{Pontecorvo:1957cp, Pontecorvo:1967fh, Maki:1962mu}, several experimental anomalies~\cite{SAGE:2009eeu, Kaether:2010ag, Barinov:2022wfh, Barinov:2021asz, Mention:2011rk, Mueller:2011nm, Huber:2011wv, Serebrov:2020kmd, Aguilar:2001ty, Aguilar-Arevalo:2013pmq, Aguilar-Arevalo:2020nvw} can possibly be explained by a hypothetical sterile neutrino with a mass at the eV scale~\cite{Pontecorvo:1967fh, Abazajian:2012ys}. 
The SAGE~\cite{SAGE:2009eeu} and GALLEX~\cite{Kaether:2010ag} experiments, and more recently, the BEST~\cite{Barinov:2022wfh, Barinov:2021asz} experiment, have observed lower than expected \nue\ rates from radioactive sources, which is known as the gallium anomaly.
Reactor neutrino experiments have measured lower $\bar\nu_e$ rates~\cite{Mention:2011rk} than the expectation based on reactor anti-neutrino flux calculations~\cite{Mueller:2011nm, Huber:2011wv}. This observation is referred to as the reactor anomaly.
An oscillation signal in the reactor $\bar{\nu}_e$ energy spectrum over distances of a few meters was reported by the \mbox{Neutrino-4}~\cite{Serebrov:2020kmd} collaboration. In addition to these observed $\overset{\brabarb}{\nu}_e$ deficits, excesses of $\overset{\brabarb}{\nu}_e$-like events were also observed in some $\overset{\brabarb}{\nu}_{\mu}$ dominated accelerator neutrino experiments. The LSND collaboration~\cite{Aguilar:2001ty} observed an anomalous excess of $\bar\nu_e$-like events, and the MiniBooNE collaboration~\cite{Aguilar-Arevalo:2013pmq, Aguilar-Arevalo:2020nvw} observed an excess of low-energy electron-like events.

These anomalies are in strong tension with other experimental results within the 3(active) + 1(sterile) oscillation framework as seen in a global fit of the data~\cite{Giunti:2019aiy}. In addition, recent experimental measurements~\cite{DayaBay:2017jkb, DayaBay:2021dqj} and improvements of the reactor anti-neutrino flux calculation~\cite{Kopeikin:2021ugh, Giunti:2021kab} lead to a plausible resolution of the reactor anti-neutrino anomaly. The \mbox{Neutrino-4} anomaly is largely excluded by the results from other very short baseline reactor neutrino experiments, for example, PROSPECT~\cite{PROSPECT:2020sxr}, STEREO~\cite{STEREO:2019ztb}, DANSS~\cite{Danilov:2021oop}, NEOS~\cite{RENO:2020hva}, although it is consistent with the gallium anomaly.

The MicroBooNE collaboration has recently reported a first set of searches related to the MiniBooNE low-energy excess, targeting multiple final-state topologies of the charged-current (CC) \nue\ interactions~\cite{MicroBooNE:2021ktl, MicroBooNE:2021bcu, MicroBooNE:2021pld, MicroBooNE:2021nxr} and the neutral-current (NC) $\Delta$ resonance decay that produces a single photon in the final state~\cite{MicroBooNE:2021zai}. The MicroBooNE detector~\cite{Acciarri:2016smi} has a similar location and is exposed to the same booster neutrino beam (BNB)~\cite{Stancu:2001cpa} as the MiniBooNE detector. Utilizing the liquid argon time projection chamber (LArTPC) technology that can provide good $e/\gamma$ separation, MicroBooNE has achieved high-performance \nue\ selections and observes no evidence of a \nue\ excess~\cite{MicroBooNE:2021ktl, MicroBooNE:2021bcu, MicroBooNE:2021pld, MicroBooNE:2021nxr}. These results disfavor the hypothesis that the MiniBooNE low-energy excess originates solely from an excess of $\nu_e$ interactions. Instead, one or more additional mechanisms~\cite{deGouvea:2019qre, Vergani:2021tgc, Asaadi:2017bhx, Alves:2022vgn, Bertuzzo:2018itn, Ballett:2018ynz, Abdallah:2020vgg, Abdallah:2020biq} are required to explain the MiniBooNE observations.

A light sterile neutrino would profoundly impact fundamental physics. In addition to testing models that may explain both the MicroBooNE and MiniBooNE low-energy \nue\ observations, interpreting the MicroBooNE \nue\ results in the context of a sterile neutrino can provide valuable statements beyond the conclusions already reached by the current analyses, and examine the remaining experimental anomalies that may be explained by a sterile neutrino. 
Recent phenomenological studies have examined the MicroBooNE \nue\ results in the context of a sterile neutrino hypothesis. One study~\cite{Denton:2021czb} considers a \nue\ disappearance-only hypothesis, while another~\cite{Arguelles:2021meu} considers the full $3+1$ oscillation effect. 

In this Letter, we present a new analysis testing the sterile neutrino hypothesis in a full $3+1$ oscillation framework with detailed event-level information. We use the data set from the MicroBooNE inclusive \nue\ CC measurement~\cite{MicroBooNE:2021nxr}, and compare the results to the parameter space allowed by the LSND, gallium (including BEST), and \mbox{Neutrino-4} anomalies.
We simultaneously consider short-baseline sterile-neutrino-induced \nue\ appearance and \nue\ disappearance. This treatment can lead to cancellations that result in a degeneracy when determining the oscillation parameters, which we will introduce in more detail in this paper.

The MicroBooNE detector~\cite{Acciarri:2016smi} is a 10.4\,m long, 2.6\,m wide, and 2.3\,m tall LArTPC, located on-axis of the BNB at Fermilab. It consists of about 85 metric tons of liquid argon in the TPC active volume for ionization charge detection along with an array of photomultiplier tubes~\cite{Briese:2013wua} for scintillation light detection. It sits at a distance of 468.5\,m from the target of the BNB, which uses protons with a kinetic energy of $8$\,GeV impinging on the target, producing secondary hadrons. The hadrons are mostly pions or kaons that decay in flight, producing a neutrino beam through their decay. The MicroBooNE BNB data set was collected entirely in neutrino mode and consists of a very pure \numu\ beam with a small $\bar{\nu}_{\mu}$ contamination and a \nue\ contamination of $<1\%$. 

We perform a full $3+1$ ($4\nu$) neutrino oscillation analysis, capitalizing on the seven channels of \nue\ and \numu\ selections and their statistical and systematic uncertainties from the MicroBooNE inclusive \nue\ low-energy excess search~\cite{MicroBooNE:2021nxr}. The analysis uses the BNB \mbox{Runs 1--3} data set with an exposure of 6.369$\times$10$^{20}$ protons on target (POT). In addition to the standard Monte Carlo (MC) samples for intrinsic \nue\ and \numu\ events in the BNB, a dedicated $\nu_{\mu}\rightarrow\nu_e$ oscillation sample was generated to appropriately take into account the flux and cross-section systematic uncertainties related to the \nue\ appearance events. The seven channels comprise fully contained (FC) and partially contained (PC) \nue\ CC processes, FC and PC \numu\ CC processes without final-state \pio\ mesons, FC and PC \numu\ CC processes with final-state \pio\ mesons, and a NC channel with final-state \pio\ mesons. The fully contained events are defined as those that have all reconstructed TPC activity (i.e., charge depositions) within a fiducial volume 3\,cm from the TPC boundaries.
Because there are \numu\ and \nue\ components in the BNB flux, the \nue\ appearance (from $\nu_{\mu}$), \nue\ disappearance, and \numu\ disappearance oscillation effects in the $3+1$ framework are simultaneously applied to the predicted signal and background events in \textit{all} seven channels in the oscillation fit. The \numu\ appearance effect is neglected because of the very low fraction of intrinsic \nue\ in the BNB flux.
This strategy takes full advantage of the statistics of the selected \nue\ and \numu\ events in the FC and PC channels, and at the same time maintains the capability to apply data constraints across channels through a joint fit to the seven channels, thereby reducing the systematic uncertainty in the oscillation analysis. The neutrino energy reconstruction primarily follows a calorimetric method with an energy resolution of approximately $10\text{--}15\%$ and a bias of $5\text{--}10\%$ for CC events~\cite{MicroBooNE:2021nxr}. In the reconstruction of NC events, we use this method to estimate the energy transfer with an invisible outgoing neutrino. The reconstruction of visible energy for the NC events in this analysis has a similar bias and energy resolution to the neutrino energy reconstruction of CC events.  

We use an extended $4\times4$ unitary PMNS matrix ($U$) to describe the $3+1$ neutrino mixing between the flavor and mass eigenstates. Following the common parameterization~\cite{Harari:1986xf, Giunti:2019aiy}, the elements of $U$ relevant to this Letter can be expressed as
\begin{align}
    |U_{e4}|^2 &= \mathrm{sin}^2\theta_{14}, \nonumber \\
    |U_{\mu4}|^2 &= \mathrm{cos}^2\theta_{14}~\mathrm{sin}^2\theta_{24}, \\
    |U_{s4}|^2 &= \mathrm{cos}^2\theta_{14}~\mathrm{cos}^2\theta_{24}~\mathrm{cos}^2\theta_{34}, \nonumber
\end{align}
where $s$ denotes the sterile neutrino flavor.
Given the energy range of the neutrino flux at MicroBooNE, in the parameter space with $\Delta m_{41}^{2} \gg |\Delta m_{31}^{2}|$, the short-baseline oscillation probability from $\alpha$-flavor to $\beta$-flavor neutrinos in vacuum approximates to 
\begin{equation}
    P_{\nu_\alpha\rightarrow\nu_\beta} = \delta_{\alpha\beta} + (-1)^{\delta_{\alpha\beta}}~ \mathrm{sin}^{2}2\theta_{\alpha\beta}~\mathrm{sin}^{2}\Delta_{41},
    \label{eq:FullProb}
\end{equation}
where $\delta_{\alpha\beta}$ is the Kronecker delta,
\begin{equation}
    \Delta_{41}\equiv\frac{\Delta m_{41}^{2}L}{4E}=1.267\left(\frac{\Delta m_{41}^{2}}{\mathrm{eV}^2}\right)\left(\frac{\mathrm{MeV}}{E}\right)\left(\frac{L}{\mathrm{m}}\right),
    \label{eq:DeltaIJ}
\end{equation}
and
\begin{equation}
    \mathrm{sin}^{2}2\theta_{\alpha\beta}=4|U_{\alpha4}|^2|\delta_{\alpha\beta}-|U_{\beta4}|^2|.
\end{equation}
We define $\theta_{\alpha\beta}$ as the effective mixing angles, which can be expressed as 
\begin{align}
    \mathrm{sin}^{2}2\theta_{ee} & = \mathrm{sin}^{2}2\theta_{14}, \nonumber \\
    \mathrm{sin}^{2}2\theta_{\mu e} & = \mathrm{sin}^{2}2\theta_{14}~\mathrm{sin}^{2}\theta_{24}, \nonumber \\
    \mathrm{sin}^{2}2\theta_{\mu\mu} & = 4\mathrm{cos}^2\theta_{14}\mathrm{sin}^2\theta_{24}(1-\mathrm{cos}^2\theta_{14}\mathrm{sin}^2\theta_{24}), \\
    \mathrm{sin}^{2}2\theta_{es} & = \mathrm{sin}^{2}2\theta_{14}~\mathrm{cos}^2\theta_{24}~\mathrm{cos}^2\theta_{34}, \nonumber \\
    \mathrm{sin}^{2}2\theta_{\mu s} & = \mathrm{cos}^{4}\theta_{14}~\mathrm{sin}^22\theta_{24}~\mathrm{cos}^2\theta_{34}. \nonumber
\end{align}
Ignoring the oscillation effect in the negligible neutrino background outside of the detector cryostat, for the other CC and NC signal or background events in all seven channels, we use $\mathrm{sin}^{2}2\theta_{ee}$ and $\mathrm{sin}^{2}2\theta_{\mu e}$ to predict the \nue\ CC energy spectrum, $\mathrm{sin}^{2}2\theta_{\mu\mu}$ to predict the \numu\ CC energy spectrum, and $\mathrm{sin}^{2}2\theta_{es}$ and $\mathrm{sin}^{2}2\theta_{\mu s}$ to predict the NC energy spectrum.
We fix $\theta_{34}$ to 0 ($\mathrm{cos}^2\theta_{34}=1$) since it has a negligible impact in this analysis given the current contribution of the NC events in the seven channels. The NC events are mainly used to constrain the NC \pio\ background in the \nue\ CC channels and the NC event disappearance can be probed in the future with a more inclusive NC selection. As a result, the three oscillation parameters $\Delta m_{41}^{2}$, $\mathrm{sin}^{2}\theta_{14}$, and $\mathrm{sin}^2\theta_{24}$ are free to vary in the fit.

It is important to note that in an oscillation analysis such as this one, performed in a \numu-dominated beam with a non-negligible intrinsic \nue\ component, the effects of \nue\ disappearance and appearance can lead to a cancellation effect on the impact on the expected event rates. Equation~\ref{eq:cancellation} demonstrates this quantitatively,
\begin{equation}
\begin{aligned}
   N_{\nu_e}(E_{\nu}) &= T_{\nu_e}(E_{\nu})[ 1 + (R(E_{\nu})\cdot{\rm sin}^2\theta_{24}-1) \\
     &\cdot {\rm sin}^22\theta_{14}\cdot{\rm sin}^2\Delta_{41}(E_{\nu})],
\end{aligned}
\label{eq:cancellation}
\end{equation}
where $T_{\nu_e}$ is the number of intrinsic \nue\ in the flux, and $R$ is the ratio between the number of intrinsic \numu\ and \nue\ for a given true neutrino energy $E_{\nu}$.
When sin$^2\theta_{24}$ approaches the inverse of the average value of $R(E_{\nu})$ in the BNB, i.e., $1/R\approx0.005$, the \nue\ appearance and \nue\ disappearance contributions mostly cancel leading to a diminished oscillation effect in the \nue\ channels, independent of the values of $\Delta m_{41}^{2}$ and $\mathrm{sin}^{2}\theta_{14}$. This results in a decreased sensitivity to sterile neutrino oscillations in this specific parameter space, which was not fully considered in some experimental results~\cite{Aguilar:2001ty, Aguilar-Arevalo:2013pmq, Aguilar-Arevalo:2020nvw}.

\begin{figure}[h!]
\centering
 \begin{overpic}[width=0.48\textwidth]{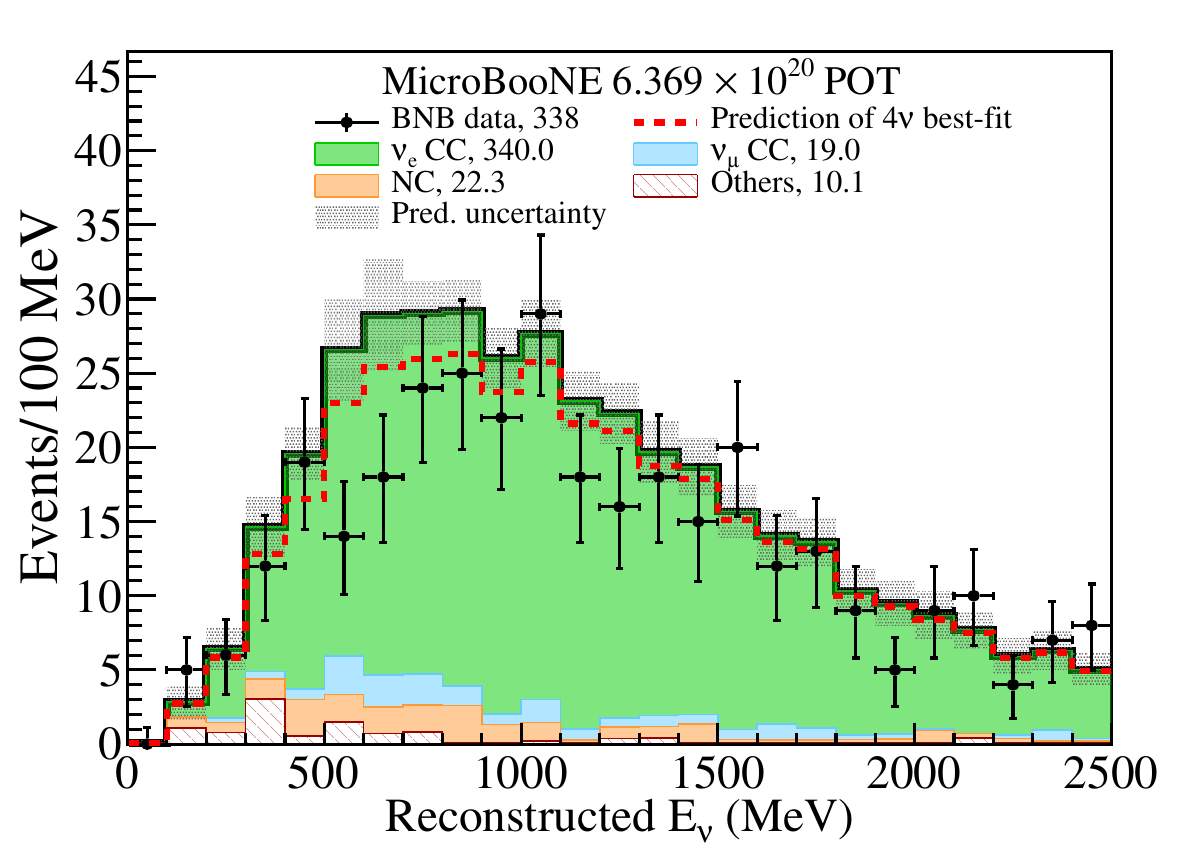}
   \put(475,-20){(a)}
  \end{overpic} 
 
 \vspace{6pt}
 
 \begin{overpic}[width=0.48\textwidth]{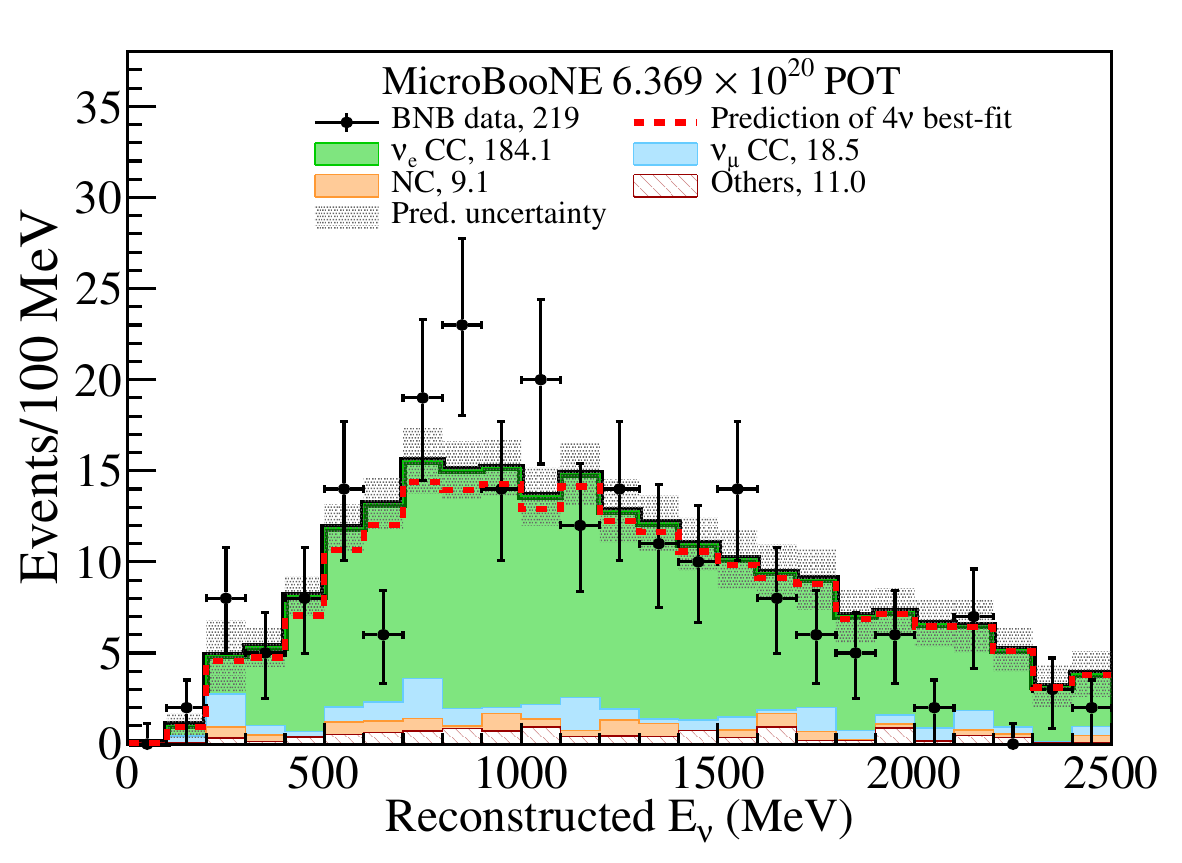}
   \put(475,-20){(b)}
  \end{overpic} 
\caption{Reconstructed neutrino energy of (a) fully contained \nue\ $\mathrm{CC}$ and (b) partially contained \nue\ $\mathrm{CC}$ events. The data points are shown with statistical error bars. The MC predictions of the $3\nu$ hypothesis for \nue\ CC events (green) and different types of backgrounds are shown in the stack of histograms. The category ``Others'' corresponds to the background events originating from either beam neutrino interactions outside the fiducial volume or cosmic-ray muons. The dashed red histogram represents the MC prediction of the $4\nu$ best-fit with $\mathrm{\Delta m^{2}_{41}=1.295\,eV^{2},\ sin^{2}\theta_{14}=0.936}$ ($\mathrm{sin^{2}2\theta_{ee}=0.240}$), and $\mathrm{sin^{2}\theta_{24}=0}$ ($\mathrm{sin^{2}2\theta_{\mu e(\mu \mu)}=0}$). The MC predictions and shaded error bands correspond to the central values and systematic uncertainties for each energy bin with constraints (Sec. VI A in Ref.~\cite{MicroBooNE:2021nxr}) from the \numu\ CC and \pio\ channels as used in the joint fit to the seven channels.}
\label{fig:nueCC_spectra}
\end{figure}

The test statistic used in the oscillation fit is the Combined-Neyman-Pearson (CNP) $\chi^2$~\cite{Ji:2019yca}
\begin{equation}
    \chi^2 = \left(M-P\right)^T\cdot \left(\mathrm{Cov}_{\rm stat}+\mathrm{Cov}_{\rm sys}\right)^{-1} \cdot \left(M-P\right),
    \label{eq:eqchi2}
\end{equation}
where $M$ and $P$ are vectors of the measurements and the predictions for the seven channels, respectively, $\mathrm{Cov}_{\rm stat}$ is the CNP-format statistical uncertainty covariance matrix corresponding to $3/\left(1/M_i + 2/P_i\right)$ for the $i$th bin, and $\mathrm{Cov}_{\rm sys}$ is the covariance matrix of the full systematic uncertainty. The systematic uncertainties are estimated from (i) the neutrino flux prediction of the BNB~\cite{AguilarArevalo:2008yp}, (ii) $\nu$-argon cross section modeling from the \textsc{Genie} event generator~\cite{GENIE:2021npt, uboone_genie_tune}, (iii) final-state hadron-argon interactions in the \textsc{Geant4} simulation~\cite{Agostinelli:2002hh, Calcutt:2021zck}, (iv) residual discrepancies in detector response after calibrations~\cite{Adams:2019qrr, MicroBooNE:2019efx, Abratenko:2020bbx, MicroBooNE:2021roa}, and (v) finite statistics of the MC samples used for central value predictions. An additional uncertainty is conservatively determined for the events that originate from the neutrino interactions outside the LArTPC cryostat.
The covariance matrices $\mathrm{Cov}_{\rm stat}$ and $\mathrm{Cov}_{\rm sys}$ depend on the prediction for the central values in each energy bin and thus vary as a function of the oscillation parameters in the fit.

The data is found to agree with the 3$\nu$ (null) hypothesis within 1 standard deviation ($\sigma$) significance. The joint fit to the seven channels yields a best-fit result of $\mathrm{\Delta m^{2}_{41}=1.295\,eV^{2},\ sin^{2}\theta_{14}=0.936}$, and $\mathrm{sin^{2}\theta_{24}=0}$ with a $\chi^2$ of 86.62 for 179 degrees of freedom. The best-fit values give $\mathrm{sin^{2}2\theta_{ee}}=0.240$ and $\mathrm{sin^{2}2\theta_{\mu e(\mu\mu)}}=0$, and the corresponding predicted \nue\ energy spectra are shown in Fig.~\ref{fig:nueCC_spectra}. The energy distributions of the other channels can be found in the supplemental material~\cite{supplemental}. In this oscillation fit, the $\chi^2$ value is largely symmetric relative to $\mathrm{sin^{2}\theta_{14}=0.5}$ because the dominant oscillation effects from \nue\ appearance and \nue\ disappearance depend on $\mathrm{sin^{2}2\theta_{14}}$. The best-fit slightly prefers $\mathrm{sin^{2}\theta_{14}=0.936}$ to $\mathrm{sin^{2}\theta_{14}=0.064}$. We obtain a $\mathrm{\Delta\chi^2_{data}=\chi^2_{null, 3\nu}-\chi^2_{min,4\nu}}=2.53$ with 3 degrees of freedom, corresponding to a $p$-value of 0.426 following the Feldman-Cousins (F-C) procedure~\cite{PhysRevD.57.3873}. The supplemental material~\cite{supplemental} presents the F-C $\Delta\chi^2$ distribution corresponding to the null hypothesis. It also provides the values of $\mathrm{\Delta\chi^2_{data}=\chi^2_{4\nu}-\chi^2_{min,4\nu}}$ for each $4\nu$ hypothesis in an $80\times60\times60$ three-dimensional grid of the oscillation parameters spanning over 0.01--100\,eV$^2$ in $\Delta m_{41}^{2}$, 0.0001--1.0 in $\mathrm{sin}^{2}\theta_{14}$, and 0.0001--1.0 in $\mathrm{sin}^2\theta_{24}$ on a logarithmic scale.

Since the data is found to be consistent with the 3$\nu$ hypothesis, exclusion limits are calculated using the frequentist-motivated $\mathrm{CL_{s}}$ method~\cite{Read:2002hq}, which is commonly used for the discovery or exclusion limits in neutrino oscillation analyses~\cite{MINOS:2020iqj, PROSPECT:2020sxr, STEREO:2019ztb, Danilov:2021oop}. The $\mathrm{CL_{s}}$ test statistic is based on $\Delta \chi^2_{\rm CL_s}=\chi^2_{4\nu}-\chi^2_{3\nu}$, which compares the null $3\nu$ hypothesis and an alternative $4\nu$ hypothesis. It is defined by 
\begin{equation}
    \mathrm{CL_{s}} = \frac{1-p_{4\nu}}{1-p_{3\nu}}, 
\end{equation}
where $p_{4\nu}$ ($p_{3\nu}$) is the $p$-value of $\Delta\chi^2_{\rm CL_s, data}$ assuming the $4\nu$ (null $3\nu$) hypothesis is true. The $p$-value is determined in a frequentist approach by throwing pseudo-experiments following the corresponding full covariance matrix assuming a hypothesis is true. The region with $\mathrm{CL}_{s}\leq1-\alpha$ is excluded at the confidence level (CL) of $\alpha$.

\begin{figure*}[ht!]
\centering
 \begin{overpic}[width=0.46\textwidth]{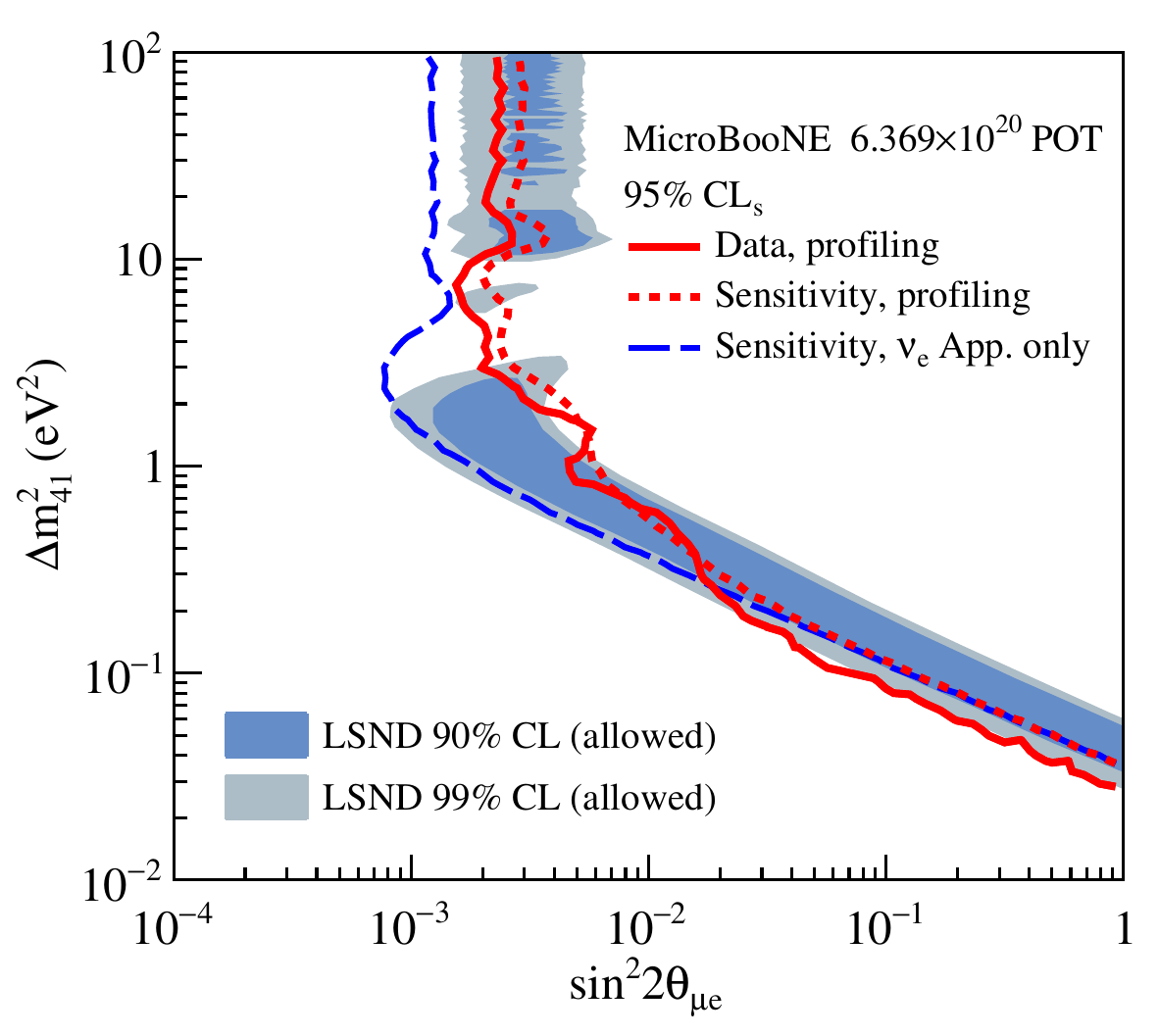}
   \put(475,-25){(a)}
  \end{overpic}
 \begin{overpic}[width=0.46\textwidth]{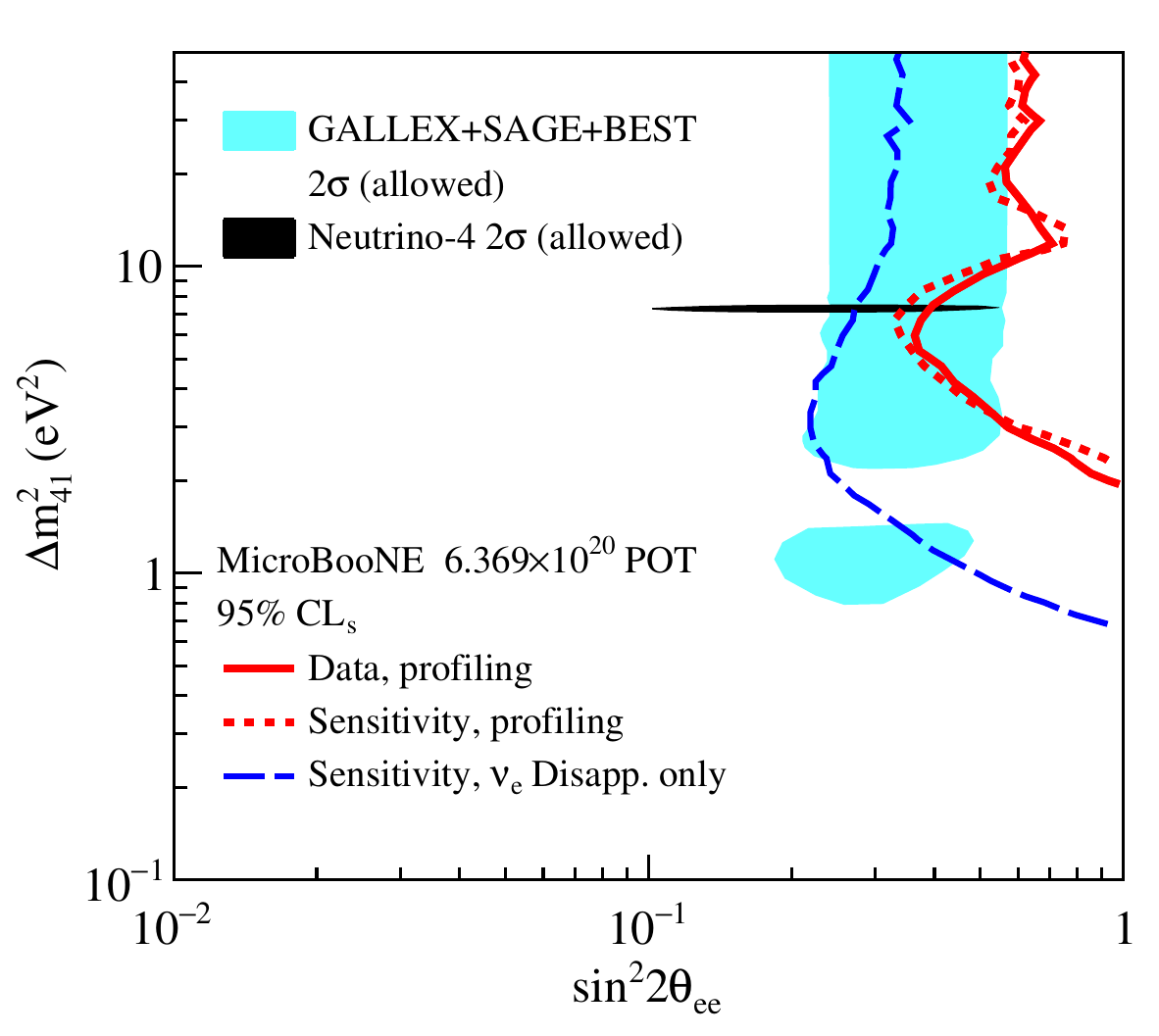}
   \put(475,-25){(b)}
  \end{overpic}
\caption{MicroBooNE $\mathrm{CL_{s}}$ exclusion contours at the $95\%$ CL in the plane of $\mathrm{\Delta m^{2}_{41}}$ and (a) $\mathrm{sin^{2}2\theta_{\mu e}}$ or (b) $\mathrm{sin^{2}2\theta_{ee}}$. The red solid (dashed) curve represents the MicroBooNE $95\%$ $\mathrm{CL_{s}}$ data exclusion (Asimov sensitivity) limits after profiling over the mixing angle $\mathrm{sin^{2}\theta_{24}}$. The blue long-dashed curve represents the MicroBooNE $95\%$ $\mathrm{CL_{s}}$ Asimov sensitivity in the scenario of (a) \nue\ appearance-only or (b) \nue\ disappearance-only as opposed to the full $3+1$ oscillation result.
In (a), the LSND $90\%$ and $99\%$ CL allowed regions~\cite{Aguilar:2001ty} using the \nue\ appearance-only approximation are shown as the light blue and gray shaded areas, respectively. In (b), the cyan shaded area represents the 2$\sigma$ allowed region of the gallium anomaly from the experimental results of GALLEX, SAGE, and BEST~\cite{Barinov:2021asz}. The 2$\sigma$ allowed region of the \mbox{Neutrino-4} experiment~\cite{Serebrov:2020kmd} is also shown in (b).} \label{fig:bnb_results}
\end{figure*}

Figure~\ref{fig:bnb_results} shows the frequentist $\mathrm{CL_{s}}$ exclusion contours and sensitivities at the $95\%$ CL in the ($\mathrm{\Delta m^{2}_{41}}$, $\mathrm{sin^{2}2\theta_{\mu e}}$) plane and in the ($\mathrm{\Delta m^{2}_{41}}$, $\mathrm{sin^{2}2\theta_{ee}}$) plane.
Since there are three free oscillation parameters in the fit, the exclusion limit in any two-dimensional (2D) parameter space is obtained by profiling the third dimension. 
After profiling, the exclusion limit corresponds to the value of the third dimension that gives the minimal $\chi^2_{4\nu}$ along that dimension at each point in the 2D parameter space. This procedure is a natural choice according to Refs.~\cite{chuang2000hybrid, Sen2009hybrid, NOvA:2022wnj}.
The $\mathrm{sin^{2}\theta_{24}}$ value after profiling in this analysis is generally small, between 0 and 0.01, which is consistent with the existing experimental constraints~\cite{Giunti:2019aiy, MINOS:2017cae, IceCube:2020phf}.
All sensitivities in this Letter are calculated using the Asimov data set~\cite{Cowan:2010js} from MC simulation, corresponding to the $3\nu$ central value predictions without oscillation.

The Asimov sensitivities in the scenarios with only \nue\ appearance or only \nue\ disappearance are often quoted in the literature~\cite{Aguilar:2001ty, Aguilar-Arevalo:2013pmq, Aguilar-Arevalo:2020nvw, T2K:2014xvp, DUNE:2020fgq} as an approximation, neglecting the oscillation effects from the intrinsic \nue\ or \numu\ component in the beam. These approximations result in overly optimistic sensitivities compared to the 2D profiled results because the cancellation between \nue\ appearance and \nue\ disappearance is neglected. Our primary result, therefore, does not use this approximation, but we include data exclusion limits taking only \nue\ appearance or only \nue\ disappearance into account in the supplemental materials~\cite{supplemental} in order to compare to historical results.

The \nue\ disappearance-only case corresponds to $\mathrm{sin^{2}\theta_{24}}=0$. However, \nue\ appearance-only is a valid approximation \textit{only} when the intrinsic \nue\ disappearance effect is small compared to the \nue\ appearance effect since non-zero \nue\ appearance requires both non-zero \nue\ and \numu\ disappearances. As seen in Fig.~\ref{fig:bnb_results}(a), the \nue\ appearance-only sensitivity asymptotically converges with the 2D profiled sensitivity in the low $\mathrm{\Delta m^{2}_{41}}$ ($<0.2$\,eV$^2$) region, where the effect of \nue\ disappearance becomes negligible compared to the \nue\ appearance effect.

The LSND allowed region shown in Fig.~\ref{fig:bnb_results}(a) was calculated using the \nue\ appearance-only approximation. After considering \nue\ disappearance, it will move towards larger $\mathrm{sin^{2}2\theta_{\mu e}}$ by a small amount because the intrinsic $\bar{\nu}_e$ contribution is small compared to the observed excess of $\bar{\nu}_e$-like events in the LSND experiment. Part of the LSND-allowed region is excluded by the MicroBooNE 2D profiled result, especially in the high and low $\Delta m^{2}_{41}$ regions. 
Portions of the allowed regions of the \mbox{Neutrino-4} and gallium anomalies in Fig.~\ref{fig:bnb_results}(b) are within the MicroBooNE data exclusion limit, with part of the region between $\Delta m_{41}^{2}=3$\,eV$^2$ and 10\,eV$^2$ excluded. Other experimental constraints on the related sterile neutrino parameter space can be found in the supplemental material~\cite{supplemental}.

The MicroBooNE results shown in this Letter are predominantly limited by the impact of the degeneracy caused by \nue\ appearance and \nue\ disappearance effects on the event rate. Future analysis strategies can break this degeneracy, further improving the sensitivity reach of a 3+1 sterile neutrino search.
The degeneracy can be addressed leveraging that MicroBooNE detects neutrinos from both the BNB and NuMI beamlines.
In addition to BNB, the MicroBooNE detector is situated at 680\,m from the NuMI target and 8$^\circ$ off-axis from the NuMI beam direction, where NuMI is the neutrino beam from the main injector~\cite{Adamson:2015dkw}. It uses protons with a kinetic energy of $120$\,GeV, much higher than BNB, impinging on the target.
The ratios of the \nue\ to the \numu\ fluxes are $0.005$ and $0.04$ for the BNB and NuMI beams, respectively.
The cancellation of \nue\ disappearance and \nue\ appearance effects therefore proceeds differently for the two beams, breaking the degeneracy that would be observed in an experiment with a single beamline.
Multi-detector oscillation analyses will also help break the degeneracy in some regions because the overall cancellation effect depends on not only the $R(E_{\nu})$ term but also the oscillation term as a function of the ratio $L/E$. Such a multiple-detector strategy, as adopted by the short-baseline neutrino program (SBN)~\cite{Machado:2019oxb}, will further improve the capability to probe the sterile neutrino parameter space with substantially reduced neutrino cross-section and flux uncertainties.

In summary, the MicroBooNE BNB Run 1--3 data show no evidence of sterile neutrino oscillations and are found to be consistent with the 3$\nu$ hypothesis within 1$\sigma$ significance. The current exclusion contours, corresponding to a BNB exposure of 6.369$\times$10$^{20}$ POT, allow for a test of part of the sterile neutrino parameter space suggested by other experimental anomalies. 
This result provides the first constraints, competitive in the relatively high $\mathrm{\Delta m^{2}_{41}}$ region, on the eV-scale sterile neutrino parameter space measured in a LArTPC detector from an accelerator neutrino source.
This work paves the way for future neutrino oscillation searches with LArTPCs in the SBN and DUNE~\cite{DUNE:2020jqi} experiments. An upcoming search for sterile neutrino oscillations at MicroBooNE combining the BNB and NuMI data will improve upon the current result by breaking the parameter degeneracy in some regions and by using data from two different beamlines.

This document was prepared by the MicroBooNE collaboration using the resources of the Fermi National Accelerator Laboratory (Fermilab), a
U.S. Department of Energy, Office of Science, HEP User Facility. Fermilab is managed by Fermi Research Alliance, LLC (FRA), acting under Contract No. DE-AC02-07CH11359.  MicroBooNE is supported by the following: the U.S. Department of Energy, Office of Science, Offices of High Energy Physics and Nuclear Physics; the U.S. National Science Foundation; the Swiss National Science Foundation; the Science and Technology Facilities Council (STFC), part of the United Kingdom Research and Innovation; the Royal Society (United Kingdom); and the UK Research and Innovation (UKRI) Future Leaders Fellowship. Additional support for the laser calibration system and cosmic ray tagger was provided by the Albert Einstein Center for Fundamental Physics, Bern, Switzerland. We also acknowledge the contributions of technical and scientific staff to the design, construction, and operation of the MicroBooNE detector as well as the contributions of past collaborators to the development of MicroBooNE analyses, without whom this work would not have been possible. For the purpose of open access, the authors have applied a Creative Commons Attribution (CC BY) public copyright license to 
any Author Accepted Manuscript version arising from this submission.

\bibliography{uB_sterile_prl}

\end{document}


\title{Supplemental Materials for: ``First constraints on light sterile neutrino oscillations from combined appearance and disappearance searches with the MicroBooNE detector''}

\author{MicroBooNE Collaboration}
\maketitle

\begin{figure}[htp!]
  \centering
  \begin{overpic}[width=0.46\textwidth]{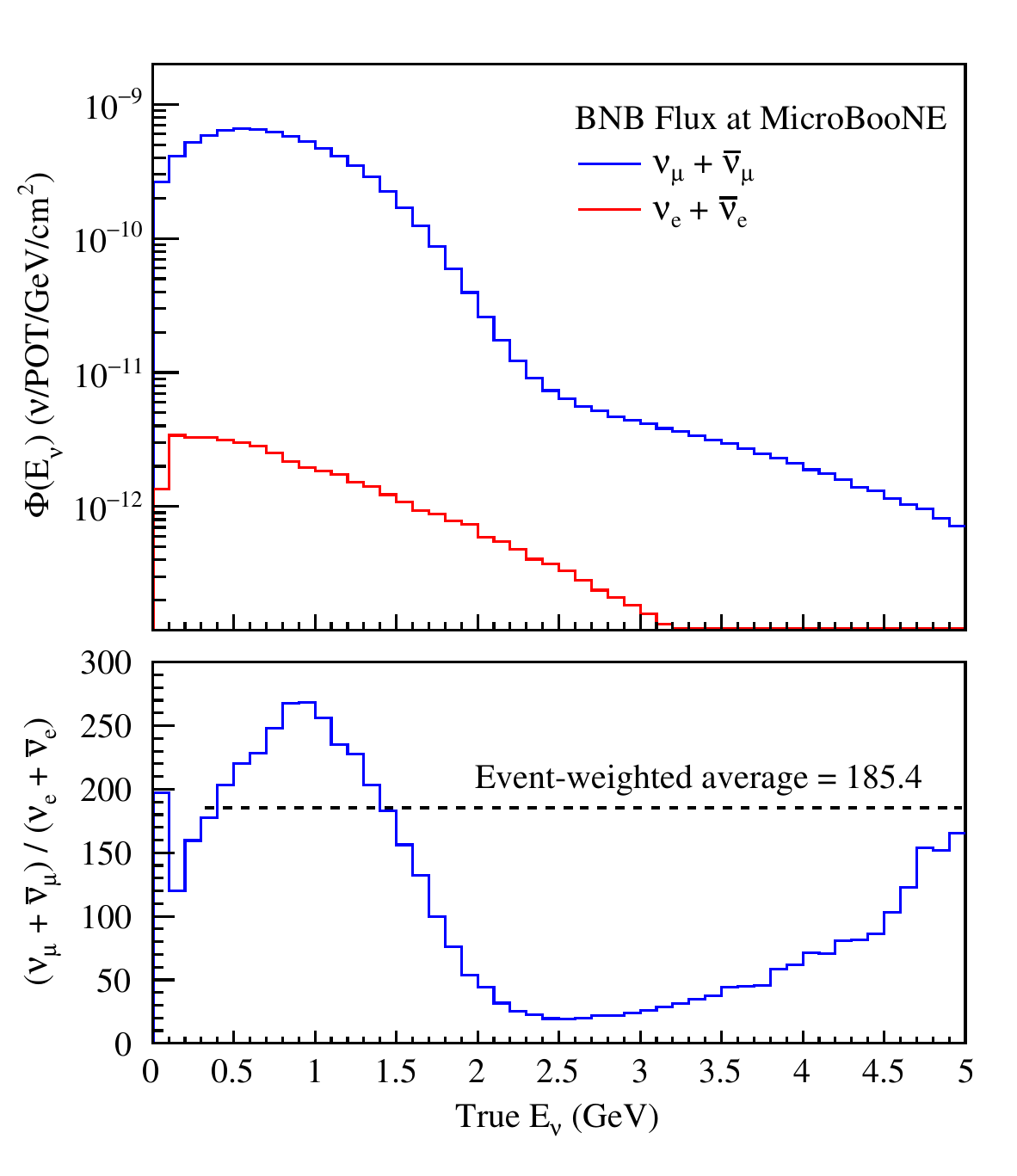}
   \put(455, -1.5){(a)}
  \end{overpic}
  \begin{overpic}[width=0.46\textwidth]{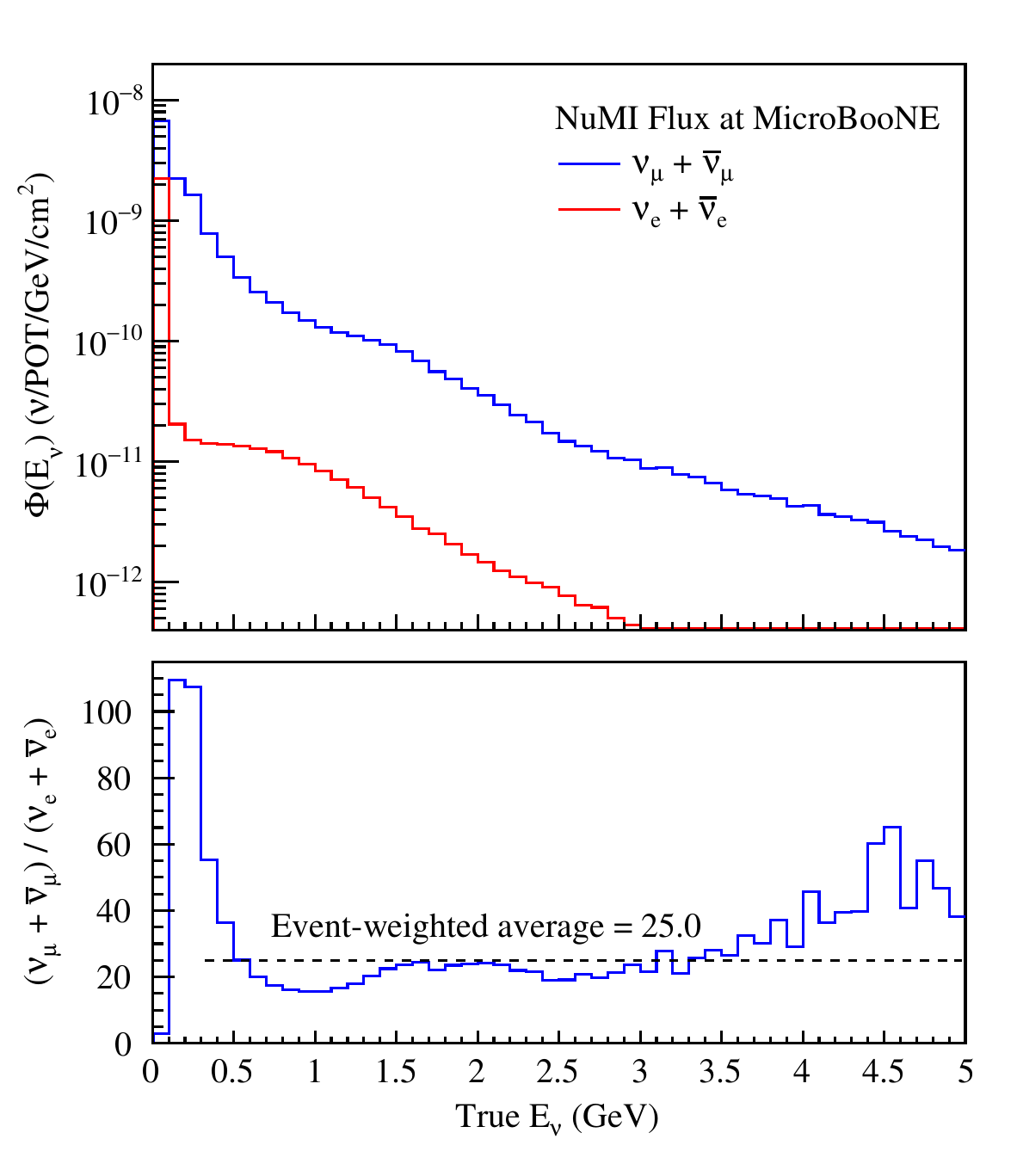}
   \put(455, -1.5){(b)}
  \end{overpic}
  \caption{Intrinsic $\nu_\mu+\bar{\nu}_\mu$ and $\nu_e+\bar{\nu}_e$ fluxes as a function of true neutrino energy for (a) BNB and (b) NuMI beams at MicroBooNE in the neutrino mode. The bottom panels show the ratios of ($\nu_\mu+\bar{\nu}_\mu$) to ($\nu_e+\bar{\nu}_e$). The event-weighted average of the ratio is calculated by using the events with neutrino energy greater than 300~MeV.}  
  \label{fig:canv_flux}
\end{figure}

Figure~\ref{fig:canv_flux} shows BNB and NuMI neutrino fluxes at MicroBooNE for $\nu_\mu+\bar{\nu}_\mu$ and $\nu_e+\bar{\nu}_e$. The ratios of ($\nu_\mu+\bar{\nu}_\mu$) to ($\nu_e+\bar{\nu}_e$) are also shown in the bottom panels. The event-weighted average values of this ratio are 185.4 and 25.0 for BNB and NuMI, respectively.

\begin{figure}[htp!]
  \centering
   \begin{overpic}[width=0.48\textwidth]{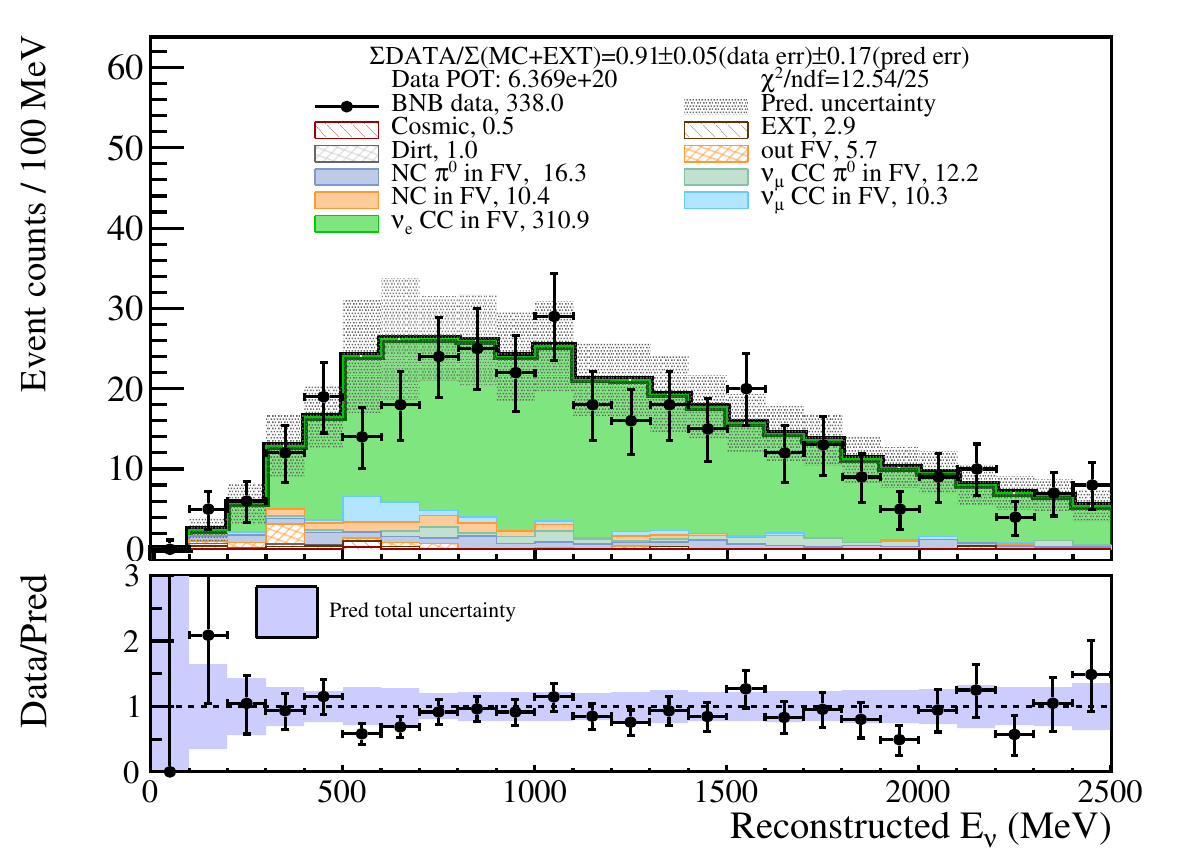}
   \put(400, -35){(a) FC $\nu_e$ CC}
   \put(620, 480){MicroBooNE}
   \end{overpic}
   \begin{overpic}[width=0.48\textwidth]{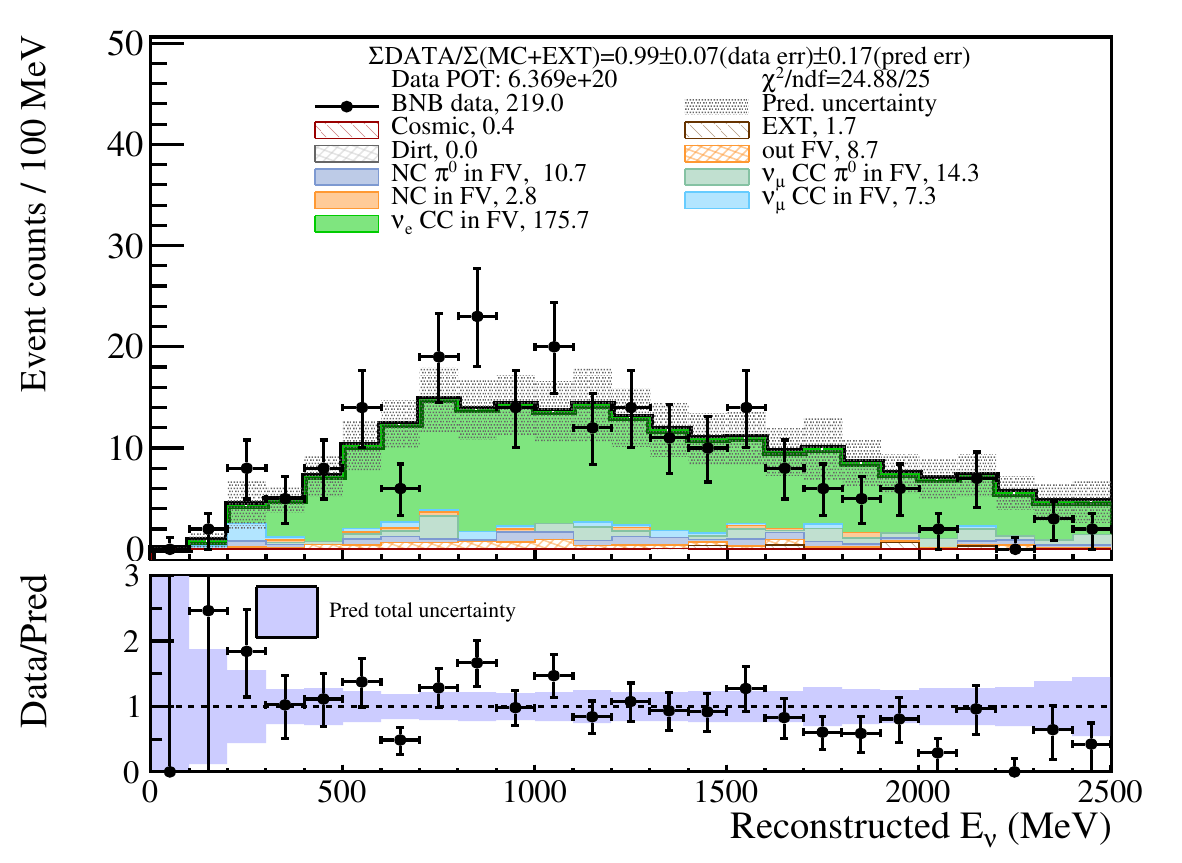}
   \put(400, -35){(b) PC $\nu_e$ CC}
   \put(620, 480){MicroBooNE}
   \end{overpic}\\
   \vspace{0.5cm}
   \begin{overpic}[width=0.48\textwidth]{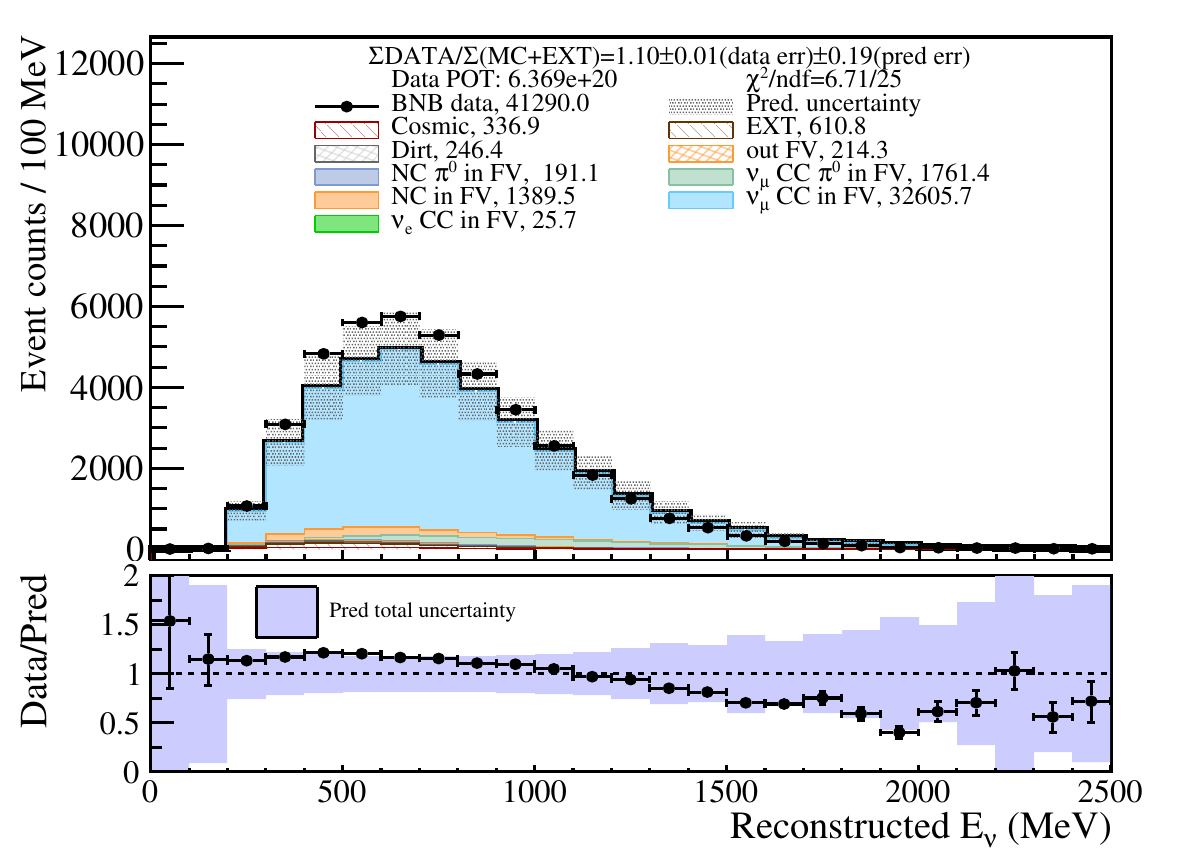}
   \put(400, -35){(c) FC $\nu_\mu$ CC}
   \put(620, 480){MicroBooNE}
   \end{overpic}
   \begin{overpic}[width=0.48\textwidth]{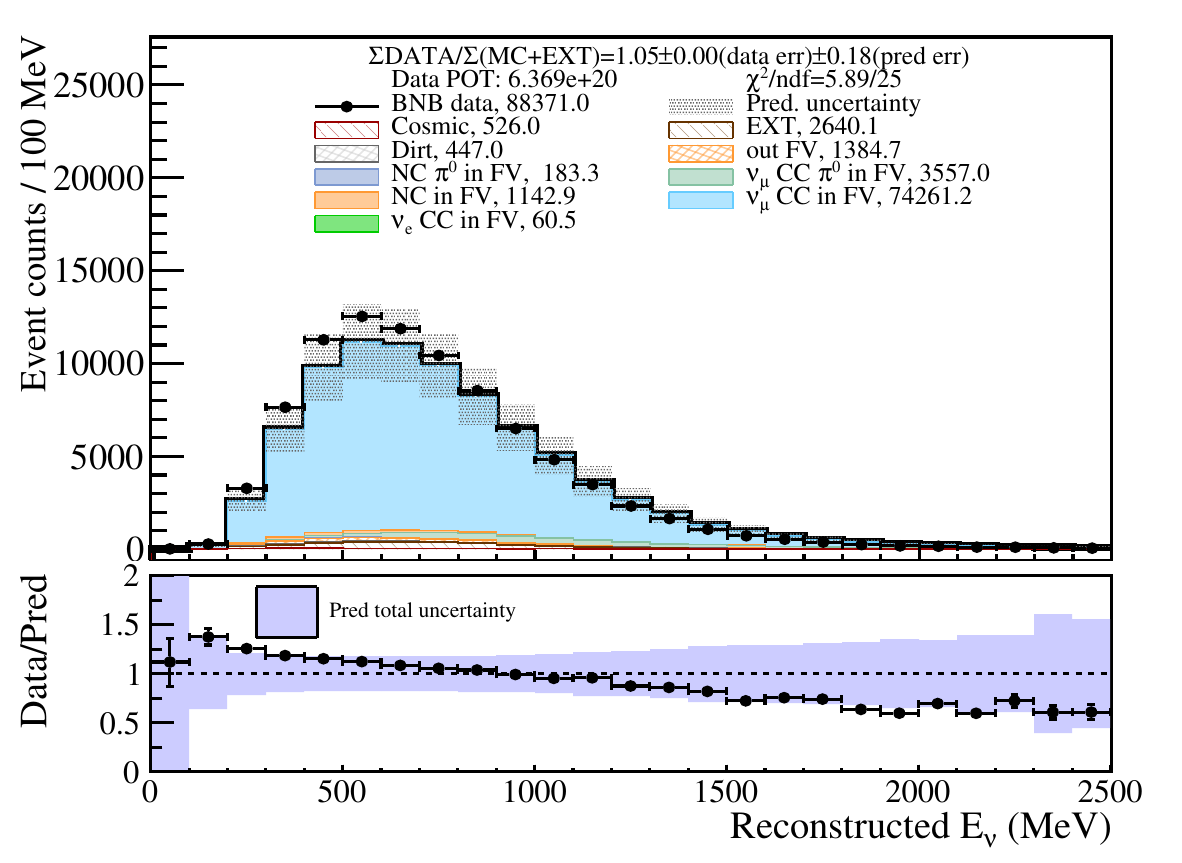}
   \put(400, -35){(d) PC $\nu_\mu$ CC}
   \put(620, 480){MicroBooNE}  
   \end{overpic}\\
   \vspace{0.5cm}
   \begin{overpic}[width=0.325\textwidth]{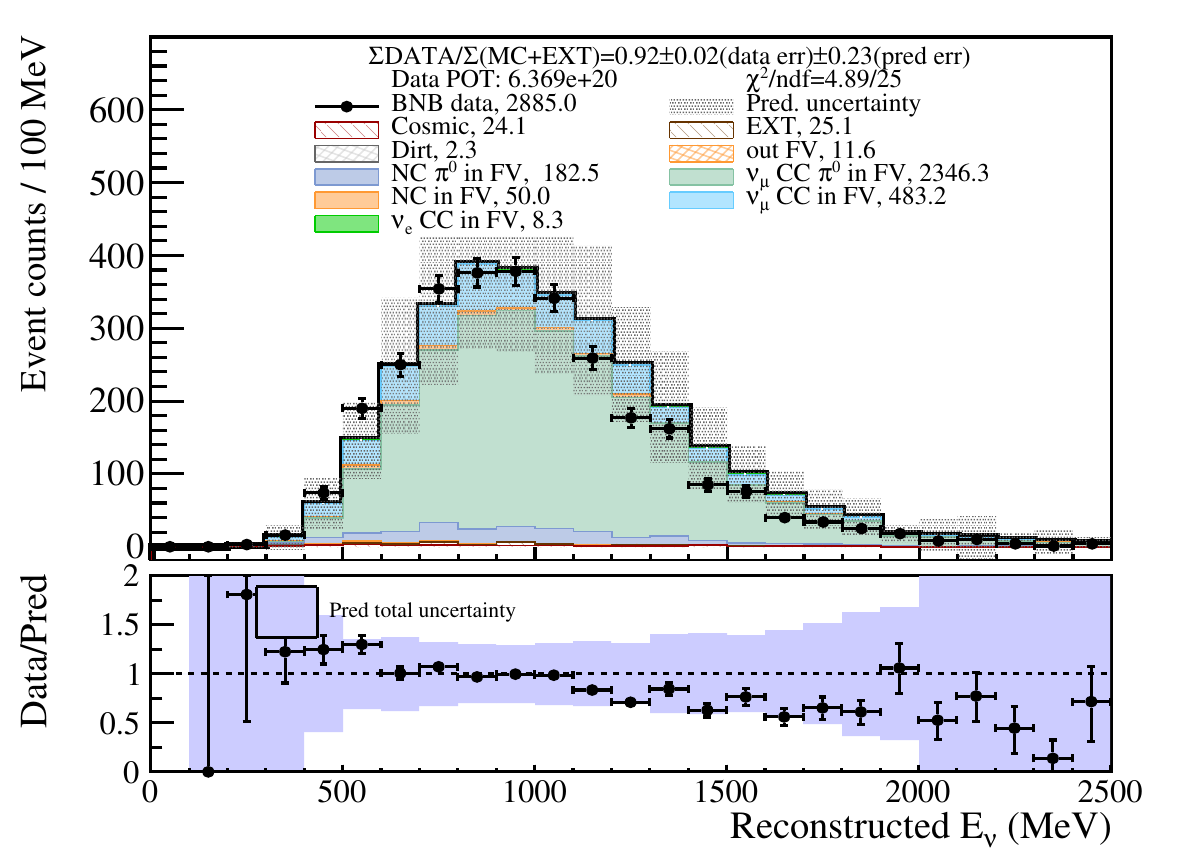}
   \put(400, -55){(e) FC CC$\pi^0$}
   \put(580, 480){MicroBooNE}
   \end{overpic}
   \begin{overpic}[width=0.325\textwidth]{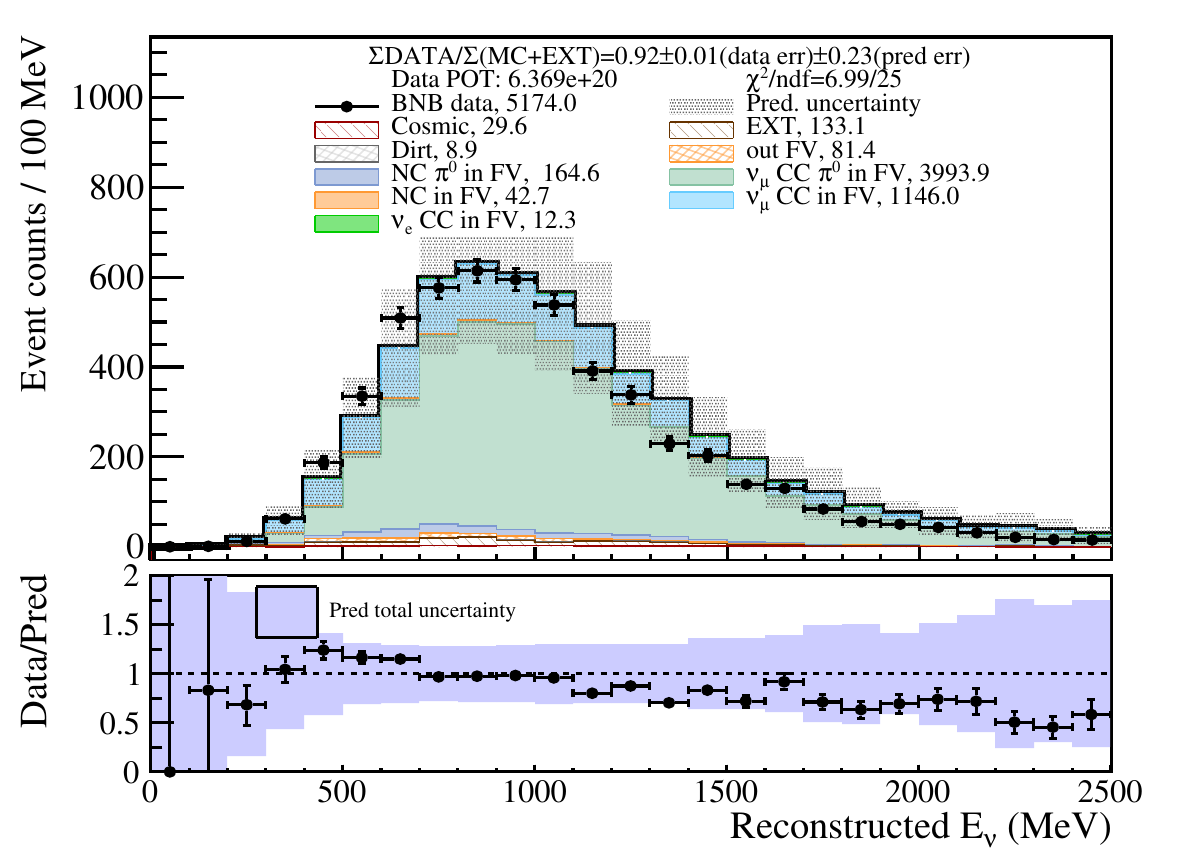}
   \put(400, -55){(f) PC CC$\pi^0$}
   \put(580, 480){MicroBooNE}
   \end{overpic}
   \begin{overpic}[width=0.325\textwidth]{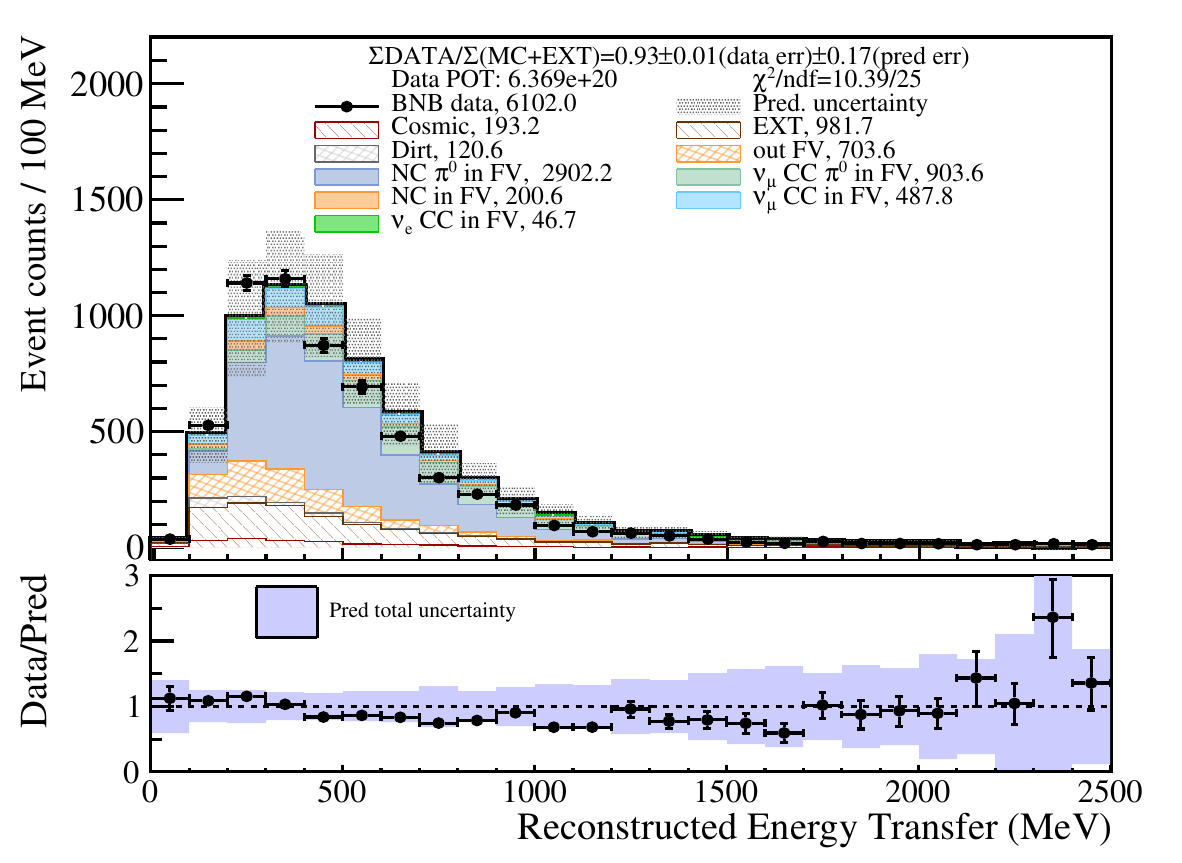}
   \put(400, -55){(g) NC$\pi^0$}
   \put(580, 480){MicroBooNE} 
   \end{overpic}\\
   \vspace{0.5cm}
  \caption{Energy spectra of the seven neutrino selection channels used in this analysis. The seven channels comprise fully contained (FC) and partially contained (PC) \nue\ CC processes (\nue\ CC), FC and PC \numu\ CC processes without final-state \pio\ mesons (\numu\ CC), FC and PC \numu\ CC processes with final-state \pio\ mesons (CC\pio), and a NC channel with final-state \pio\ mesons (NC\pio). Details of the selections and legend definitions can be found in Ref.~\cite{MicroBooNE:2021nxr}. The energy spectra of FC and PC \nue\ CC channels with constraints from the other channels under the null oscillation hypothesis, and the MC prediction of the 4$\nu$ best-fit can be found in Fig.~1 in the Letter. The 4$\nu$ best fit provides approximately the same prediction as the null oscillation hypothesis for each non-$\nu_e$ channel.}  
  \label{fig:canv_7channel}
\end{figure}

Figure~\ref{fig:canv_7channel} shows the energy spectra of the seven neutrino selection channels used in the joint fit in this sterile neutrino oscillation analysis.

\begin{figure}[htp!]
  \centering
  \includegraphics[width=0.6\textwidth]{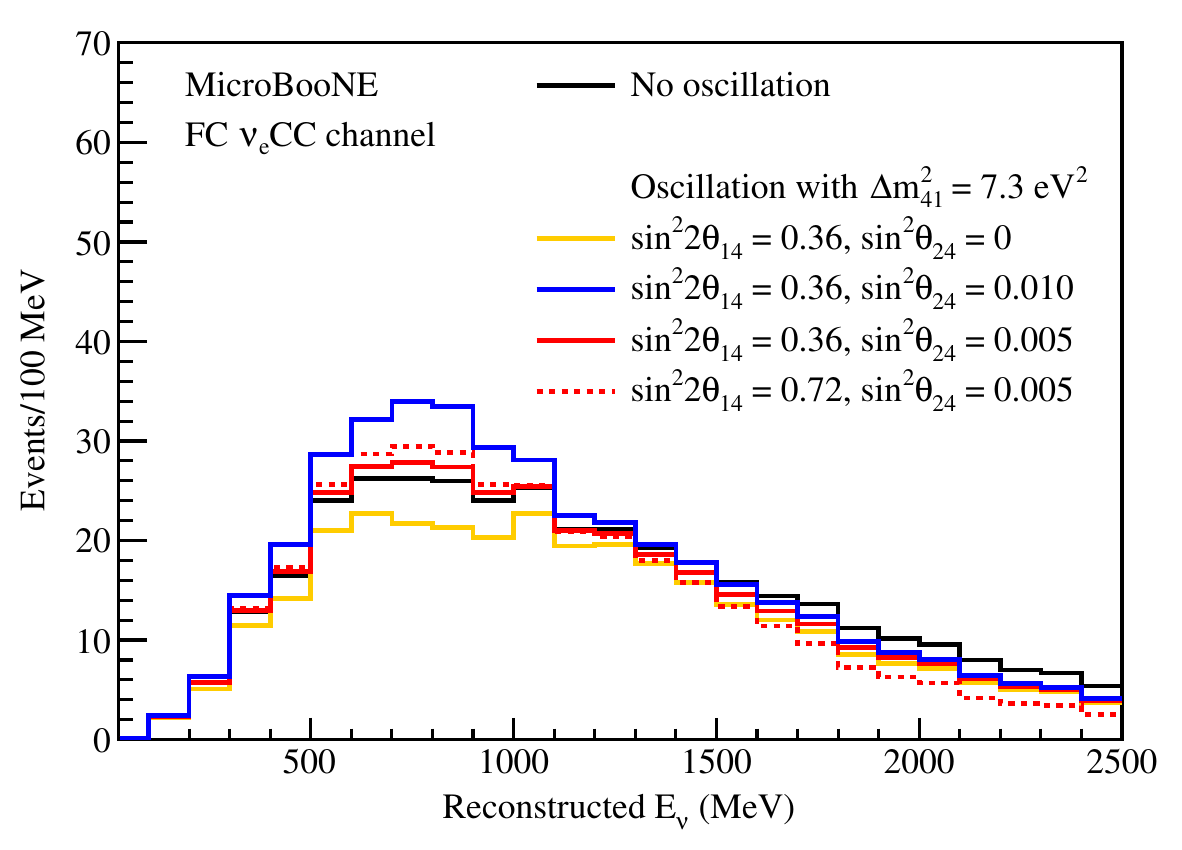}
  \caption{Energy spectra of the selected events from BNB in the fully contained (FC) $\nu_e$ CC channel for different sets of oscillation parameters.}  
  \label{fig:canv_bnb_spectra_OSC}
\end{figure}

Figure~\ref{fig:canv_bnb_spectra_OSC} shows the energy spectra of the fully contained $\nu_e$ CC events for different sets of oscillation parameters. The yellow and blue curves correspond to large oscillation effects, while the red curves show small oscillation effects due to the cancellation between $\nu_e$ disappearance and appearance when $\mathrm{sin^{2}\theta_{24}}$ approaches 0.005.

\begin{figure}[htp!]
  \centering
  \includegraphics[width=0.6\textwidth]{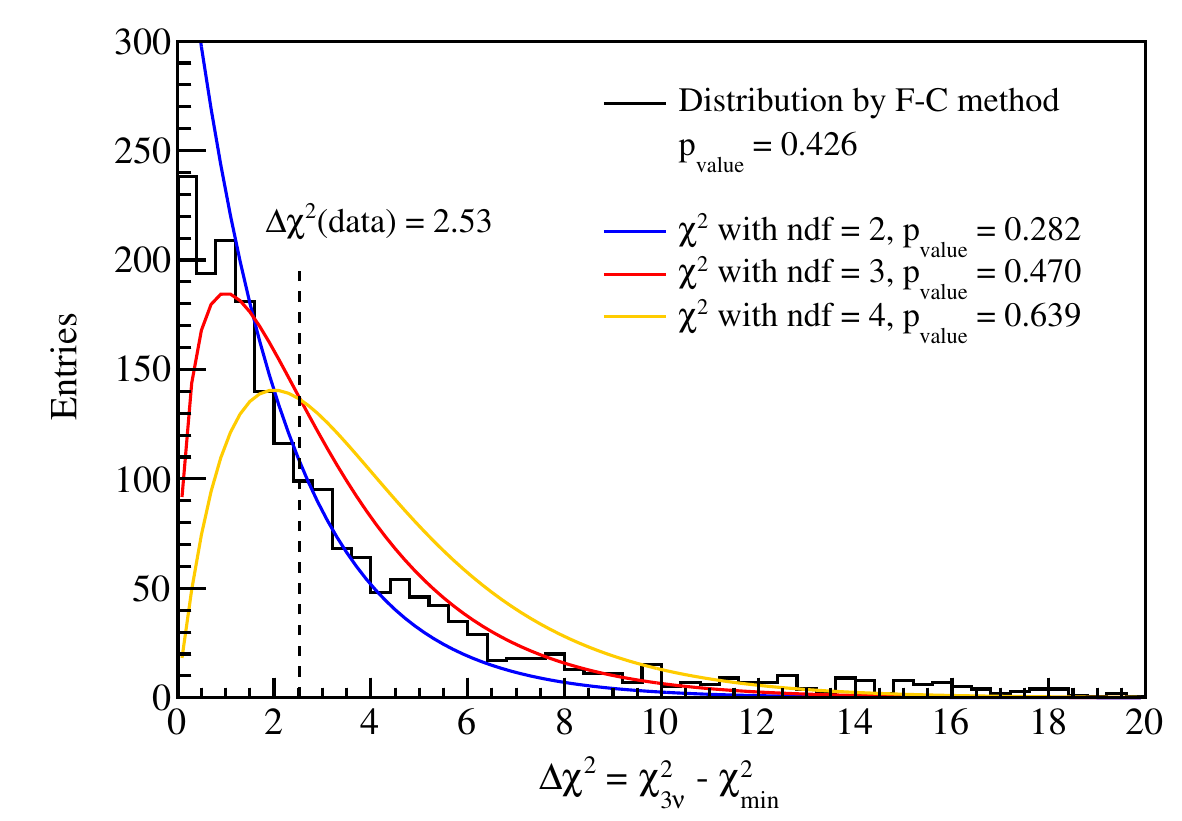}
  \caption{$\Delta\chi^{2} = \chi^2_{\mathrm{null},3\nu} - \chi^2_{\mathrm{min},4\nu}$ distribution obtained following the Feldman-Cousins (F-C) method. The figure also presents the standard $\chi^2$ distributions with number of degrees of freedom (ndf) at 2, 3 and 4.}  
  \label{fig:canv_dchi2_FC}
\end{figure}

Figure~\ref{fig:canv_dchi2_FC} shows the $\Delta\chi^{2} = \chi^2_{\mathrm{null},3\nu} - \chi^2_{\mathrm{min},4\nu}$ distribution obtained following the Feldman-Cousins method in comparison with the standard $\chi^2$ distributions with number of degrees of freedom at 2, 3 and 4.

\begin{figure*}[htp!]
 \begin{overpic}[width=0.48\textwidth]{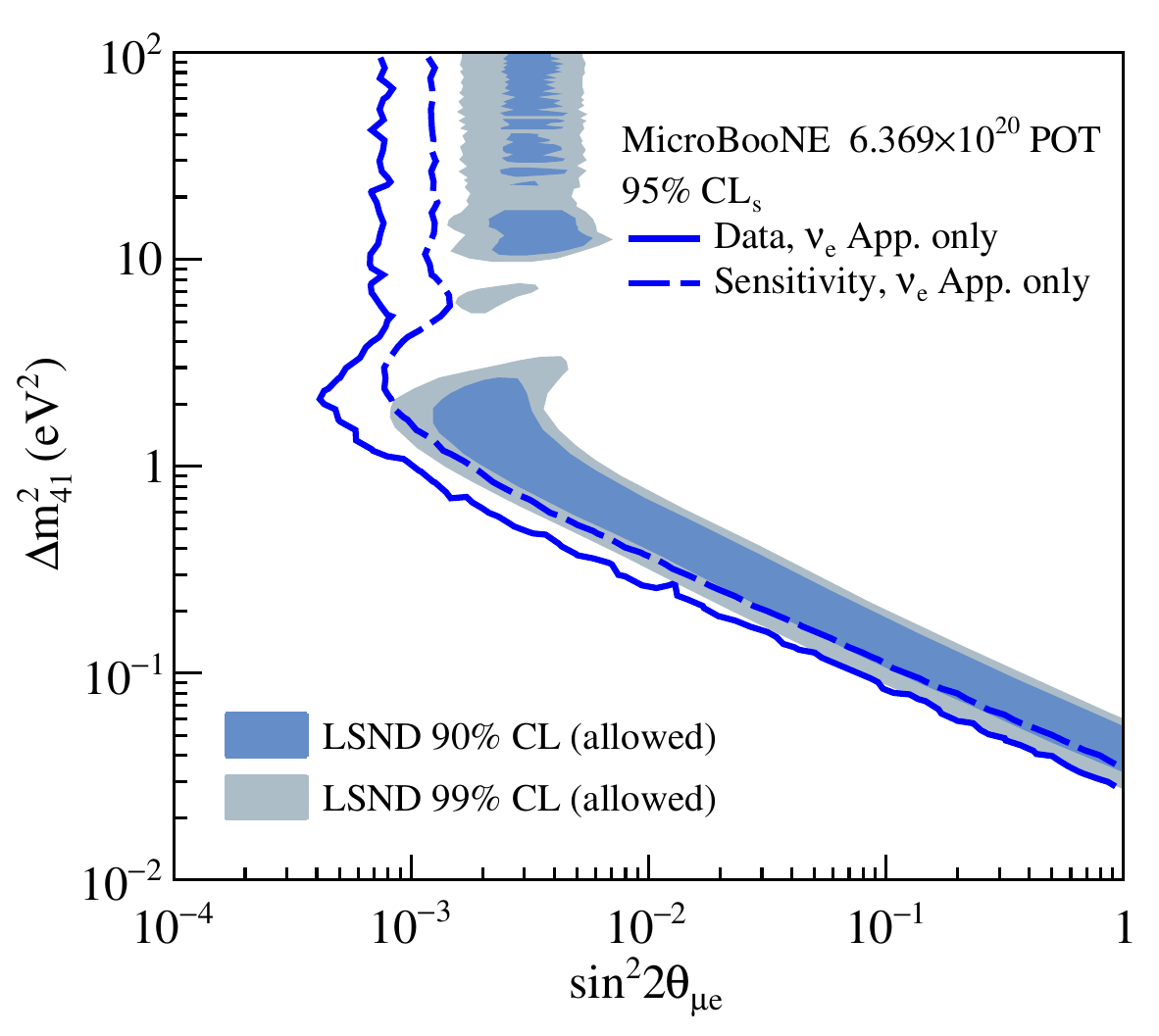}
   \put(500, -20){(a)}
  \end{overpic}
 \begin{overpic}[width=0.48\textwidth]{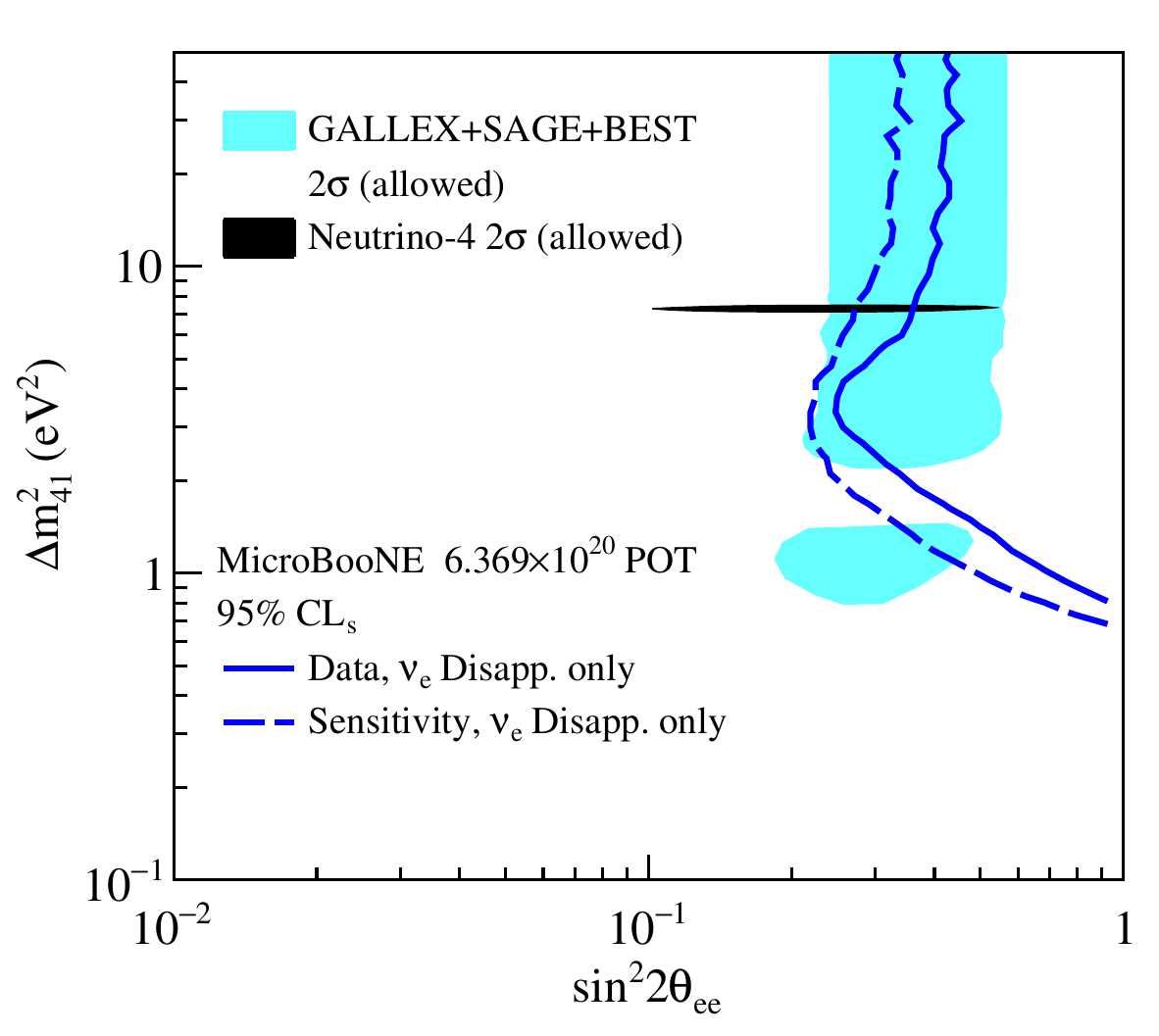}
   \put(500, -20){(b)}
  \end{overpic}
\caption{MicroBooNE $\mathrm{CL_{s}}$ sensitivity and data exclusion contours at the $95\%$ CL in the plane of $\mathrm{\Delta m^{2}_{41}}$ and (a) $\mathrm{sin^{2}2\theta_{\mu e}}$ or (b) $\mathrm{sin^{2}2\theta_{ee}}$. The blue solid (dashed) curve represents the MicroBooNE $95\%$ $\mathrm{CL_{s}}$ data exclusion (Asimov sensitivity) limits in the scenario of (a) $\nu_e$ appearance-only or (b) $\nu_e$ disappearance-only. In the left figure, the LSND $90\%$ and $99\%$ CL allowed regions~\cite{Aguilar:2001ty} using the $\nu_e$ appearance-only approximation are shown as the light blue and gray shaded areas, respectively. In the right figure, the cyan shaded area represents the 2$\sigma$ allowed region of the gallium anomaly from the experimental results of GALLEX, SAGE, and BEST~\cite{Barinov:2021asz}. The 2$\sigma$ allowed region of the Neutrino-4 experiment~\cite{Serebrov:2020kmd} is also shown.} \label{fig:bnb_results}
\end{figure*}

\begin{figure*}[htp!]
  \centering
  \includegraphics[width=0.8\textwidth]{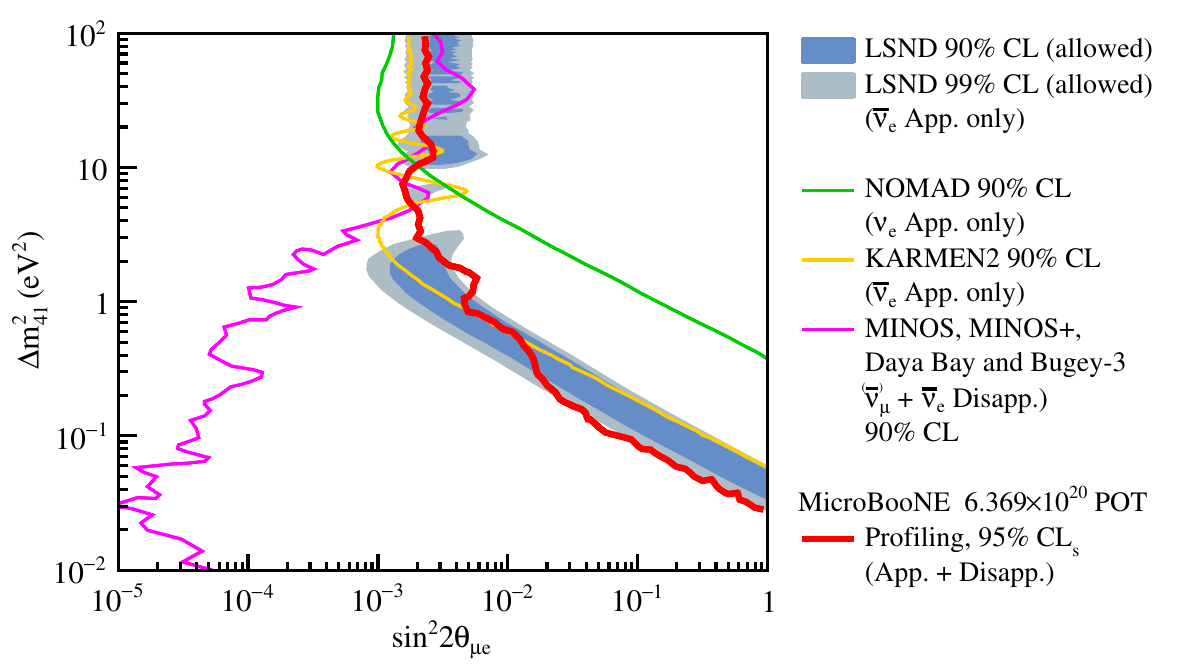}
  \caption{MicroBooNE $\mathrm{CL_{s}}$ data exclusion contour at the $95\%$ CL in the plane of $\mathrm{\Delta m^{2}_{41}}$ and $\mathrm{sin^{2}2\theta_{\mu e}}$. The figure also presents the exclusion contours at the $90\%$ CL from NOMAD~\cite{NOMAD:2003mqg}, KARMEN2~\cite{KARMEN:2002zcm}, and the combined analysis of MINOS, MINOS+, Daya Bay and Bugey-3~\cite{MINOS:2020iqj}. The LSND $90\%$ and $99\%$ CL allowed regions~\cite{Aguilar:2001ty} using the $\nu_e$ appearance-only approximation are also shown.} 
  \label{fig:canv_ue_global}
\end{figure*}

\begin{figure*}[htp!]
  \centering
  \includegraphics[width=0.8\textwidth]{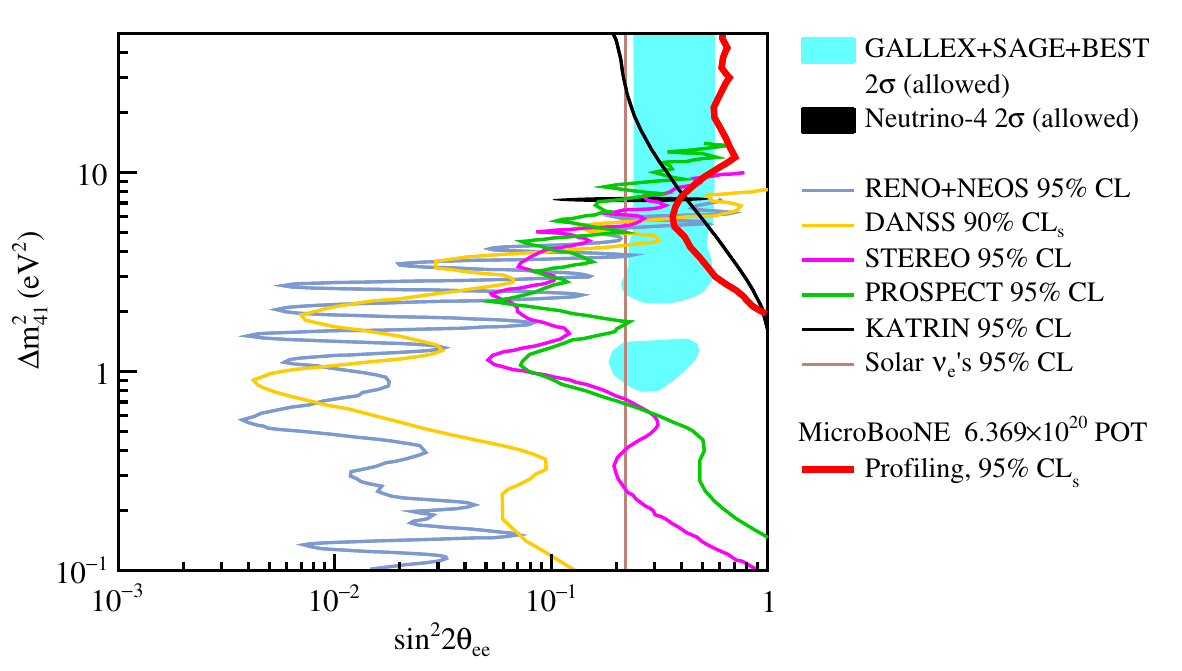}
  \caption{MicroBooNE $\mathrm{CL_{s}}$ data exclusion contour at the $95\%$ CL in the plane of $\mathrm{\Delta m^{2}_{41}}$ and $\mathrm{sin^{2}2\theta_{ee}}$. The figure also presents the exclusion contours from KATRIN~\cite{KATRIN:2022ith}, PROSPECT~\cite{PROSPECT:2020sxr}, STEREO~\cite{STEREO:2022nzk}, DANSS~\cite{Danilov:2022bss}, the combined analysis of RENO and NEOS~\cite{RENO:2020hva}, and solar $\nu_{e}$'s~\cite{Giunti:2021iti}. The 2$\sigma$ allowed region of the gallium anomaly from the experimental results of GALLEX, SAGE, and BEST~\cite{Barinov:2021asz}, and the 2$\sigma$ allowed region of the Neutrino-4 experiment~\cite{Serebrov:2020kmd} are also shown.}  
  \label{fig:canv_ee_global}
\end{figure*}

Figure~\ref{fig:bnb_results} shows MicroBooNE sensitivity and data exclusion contours at the $95\%$ $\mathrm{CL_{s}}$ in the plane of $\mathrm{\Delta m^{2}_{41}}$ and  $\mathrm{sin^{2}2\theta_{\mu e}}$ or $\mathrm{sin^{2}2\theta_{e e}}$ for $\nu_e$ appearance-only and $\nu_e$ disappearance-only scenarios.
The data and sensitivity differences in both scenarios originate from the overall deficit observed in the $\nu_e$ CC channels~\cite{MicroBooNE:2021nxr}. 
Figure~\ref{fig:canv_ue_global} and~\ref{fig:canv_ee_global} show the MicroBooNE data exclusion contours at the $95\%$ $\mathrm{CL_{s}}$ using the profiling method in the plane of $\mathrm{\Delta m^{2}_{41}}$ and $\mathrm{sin^{2}2\theta_{\mu e}}$ or $\mathrm{sin^{2}2\theta_{e e}}$ in comparison with other experimental results.


\begin{figure*}[htp!]
  \centering
  \includegraphics[width=0.55\textwidth]{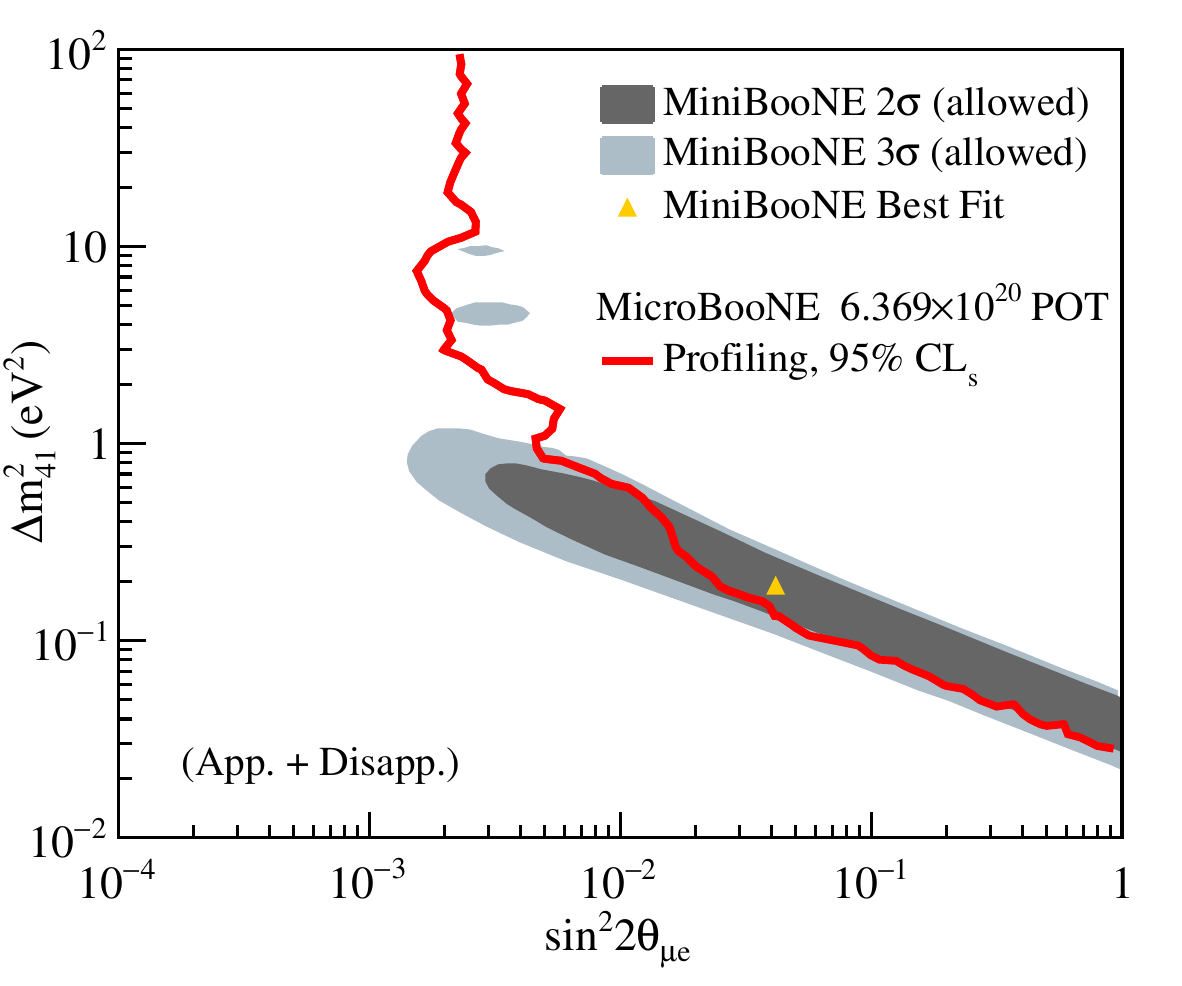}
  \caption{MicroBooNE $\mathrm{CL_{s}}$ data exclusion contour at the $95\%$ CL in the plane of $\mathrm{\Delta m^{2}_{41}}$ and $\mathrm{sin^{2}2\theta_{\mu e}}$. The MiniBooNE $2\sigma$ and $3\sigma$ CL allowed regions~\cite{MiniBooNE:2022emn}, which were calculated using the approximate test-statistic distribution from Wilks' theorem, are also shown. Both MicroBooNE and MiniBooNE results take into account all possible appearance and disappearance effects within the $3+1$ active-to-sterile neutrino oscillation framework.}  
  \label{fig:canv_ue_uB_mB}
\end{figure*}

Figure~\ref{fig:canv_ue_uB_mB} shows the MicroBooNE data exclusion contour compared with the recently updated MiniBooNE sterile neutrino oscillation result~\cite{MiniBooNE:2022emn}. Both works take into account all possible appearance and disappearance effects within the $3+1$ oscillation framework. It is worth noting that a pure $\nu_e$ excess, the assumption of the $3+1$ sterile neutrino oscillation explanation to the MiniBooNE anomaly, is disfavored by the recent MicroBooNE LEE results as mentioned in the Letter. A $2\text{--}3\sigma$ tension between the MicroBooNE and MiniBooNE results can be seen in both the $\nu_e$ low-energy excess search~\cite{MicroBooNE:2021nxr} and the sterile neutrino oscillation search as shown in this figure. 

\clearpage
\newpage


\bibliography{supplemental}{}